\documentclass[sn-aps,iicol]{sn-jnl}

\usepackage{comment}
\usepackage{siunitx}
\usepackage{physics}
\usepackage{xcolor}
\usepackage{dsfont}
\usepackage{mathtools}

\begin{document}

\title[ ]{Density-matrix renormalization group: a pedagogical introduction}

\author*[1,2,3]{G. Catarina}\email{goncalo.catarina@empa.ch}
\author[1,2]{Bruno Murta}

\affil[1]{Theory of Quantum Nanostructures Group, International Iberian Nanotechnology Laboratory (INL), 4715-330 Braga, Portugal}
\affil[2]{Centro de F\'{i}sica das Universidades do Minho e do Porto, Universidade do Minho, Campus de Gualtar, 4710-057 Braga, Portugal}
\affil[3]{Current address: nanotech@surfaces Laboratory, Empa---Swiss Federal Laboratories for Materials Science and Technology, 8600 D\"{u}bendorf, Switzerland}

\abstract{


The physical properties of a quantum many-body system can, in principle, be determined by diagonalizing the respective Hamiltonian, but the dimensions of its matrix representation scale exponentially with the number of degrees of freedom. 
Hence, only small systems that are described through simple models can be tackled via exact diagonalization. 
To overcome this limitation, numerical methods based on the renormalization group paradigm that restrict the quantum many-body problem to a manageable subspace of the exponentially large full Hilbert space have been put forth.
A striking example is the density-matrix renormalization group (DMRG), which has become the reference numerical method to obtain the low-energy properties of one-dimensional quantum systems with short-range interactions. 
Here, we provide a pedagogical introduction to DMRG, presenting both its original formulation and its modern tensor-network-based version. 
This colloquium sets itself apart from previous contributions in two ways. 
First, didactic code implementations are provided to bridge the gap between conceptual and practical understanding. 
Second, a concise and self-contained introduction to the tensor network methods employed in the modern version of DMRG is given, thus allowing the reader to effortlessly cross the deep chasm between the two formulations of DMRG without having to explore the broad literature on tensor networks. 
We expect this pedagogical review to find wide readership amongst students and researchers who are taking their first steps in numerical simulations via DMRG.

}

\maketitle

\section{Introduction}
Understanding the properties of quantum matter is one of the key challenges of the modern era~\cite{Keimer2017}.
The difficulties encountered are typically twofold.
On the one hand, there is the challenge of modelling all the interactions of a complex quantum system.
On the other hand, even when an accurate model is known, solving it is generally not an easy task.
In what follows, we will overlook the first challenge and consider only quantum systems for which we can write a model Hamiltonian.
Whether such a model is a good description of the physical system or not is thus beyond the scope of this colloquium.

Quantum problems can be divided into two classes: single-body and many-body.
In the single-body case, the model Hamiltonian does not include interactions between different quantum particles.
In other words, the quantum system can be described as if there was only one quantum particle subject to some potential.
Single-body problems are easy to solve by numerical means, as the dimension of the corresponding Hamiltonian matrix scales linearly with the number of degrees of freedom.
For instance, if we consider one electron in $N_o$ spin-degenerate molecular orbitals, we have $2N_o$ possible configurations, as the electron can have either spin-$\uparrow$ or spin-$\downarrow$ in each of the molecular orbitals.
In contrast to the single-body case, quantum many-body problems entail interactions between the different quantum particles that compose the system.
In that case, the Hamiltonian matrix must take all particles into account, which leads to an exponential growth of its dimension with the number of degrees of freedom.
Using the previous example, the basis of the most general many-body Hamiltonian should have $4^{N_o}$ terms, since every molecular orbital can be empty, doubly-occupied, or occupied by one electron with either spin-up or spin-down.
Even if we fix the number of electrons to $N_e$, we obtain $\binom{2N_o}{N_e}$ configurations, which still scales exponentially with $N_o$~\footnote{For reference, note that for $N_o = N_e = 6$, as we would have in the simplest model for a benzene molecule, there are $924$ electronic configurations, which could be encoded in $4~\textrm{kB}$ of computer memory. 
However, in the case of a slightly larger molecule such as triangulene, for which $N_o = N_e = 22$, there are roughly $2 \times 10^{12}$ configurations, which would require $8~\textrm{TB}$.}.
Hence, the exact diagonalization of quantum many-body problems is limited to small systems described by simple models.
This is known as the exponential wall problem~\cite{Kohn1999}.

In order to circumvent the exponential wall in quantum mechanics, several numerical methods, each involving a different set of approximations, have been devised.
Notable examples are the mean-field approximation, perturbation theory, the configuration interaction method~\cite{Sherrill1999}, density-functional theory~\cite{Hohenberg1964,Kohn1965,Giustino2014}, quantum Monte Carlo~\cite{Foulkes2001}, and quantum simulation~\cite{Feynman1982,Lloyd1996,Georgescu2014}, each of which having its own limitations.
Additionally, there is the density-matrix renormalization group (DMRG), introduced in 1992 by Steven R. White~\cite{White1992,White1993}.
This approach, founded on the basis of the variational principle, rapidly established itself as the reference numerical method to obtain the low-energy properties of one-dimensional (1D) quantum systems with short-range interactions~\cite{Schollwock2005}.
Importantly, a few years after its discovery, DMRG was reformulated in the language of tensor networks~\cite{Ostlund1995,Dukelsky1998,Schollwock2011}, which allowed for more efficient code implementations~\cite{Hauschild2018,Fishman2022}.
The connection between the original formulation of DMRG and its tensor-network version is by no means straightforward, as the latter involves a variational optimization of a wave function represented by a matrix product state (MPS), making no direct reference to any type of renormalization technique.

The goal of this colloquium is to present a pedagogical introduction to DMRG in both the original and the MPS-based formulations.
Our contribution should therefore add up to the vast set of DMRG reviews in the literature~\cite{Schollwock2005,Noack2005,Hallberg2006,McCulloch2007,DeChiara2008,Schollwock2011}.
By following a low-level approach and focusing on learning purposes, we aim to provide a comprehensive introduction for beginners in the field.
Bearing in mind that a thorough conceptual knowledge should be accompanied by a notion of practical implementation, we provide as supporting materials simplified and digestible code implementations in the form of documented Jupyter notebooks~\cite{Kluyver2016jupyter} to put both levels of understanding on firm footing.

The rest of this work is organized as follows.
In Section~\ref{sec:TID}, we introduce the truncated iterative diagonalization (TID).
Although this renormalization technique has been successfully applied to quantum impurity models through Wilson's numerical renormalization group~\cite{Wilson1975,Bulla2008}, we illustrate why it is not suitable for the majority of quantum problems.
Section~\ref{sec:DMRG_original} contains the original formulation of DMRG, as invented by Steven R. White~\cite{White1992,White1993}.
We first describe the infinite-system DMRG, which essentially differs from the TID by the type of truncation employed.
The truncation used in DMRG is then shown to be optimal, in the sense that it minimizes the difference between the exact and the truncated wave functions.
Importantly, we also clarify the reason that renders this truncation efficient when applied to the low-energy states of 1D quantum systems with short-range interactions.
This section ends with the introduction of the finite-system DMRG. 
In Section~\ref{sec:TN_basics}, we give a brief overview on tensor networks, addressing the minimal fundamental concepts that are required to understand how these are used in the context of DMRG.
Section~\ref{sec:finite-DMRG_TN} shows how, in the framework of tensor networks, the finite-system DMRG can be seen as an optimization routine that, provided a representation of the Hamiltonian in terms of a matrix product operator (MPO), minimizes the energy of a variational MPS wave function.
Finally, in Section~\ref{sec:conclusion}, we present our concluding remarks, mentioning relevant topics that are beyond the scope of this review.

In Supplementary Information, we make available a transparent (though not optimized) Python~\cite{Python2009} code that, for a given 1D spin model, implements the following algorithms:
i) iterative exact diagonalization, which suffers from the exponential wall problem; 
ii) TID; 
iii) infinite-system DMRG, within the original formulation.
For pedagogical purposes, this code shares the same main structure for the three methods, differing only on a few lines of code that correspond to the implementation of the truncations associated with each method.
Following the same didactic approach, we also provide a practical implementation of the finite-system DMRG algorithm in the language of tensor networks.

\section{Truncated iterative diagonalization}\label{sec:TID}

\begin{figure}
 \centering
 \includegraphics[width=\columnwidth]{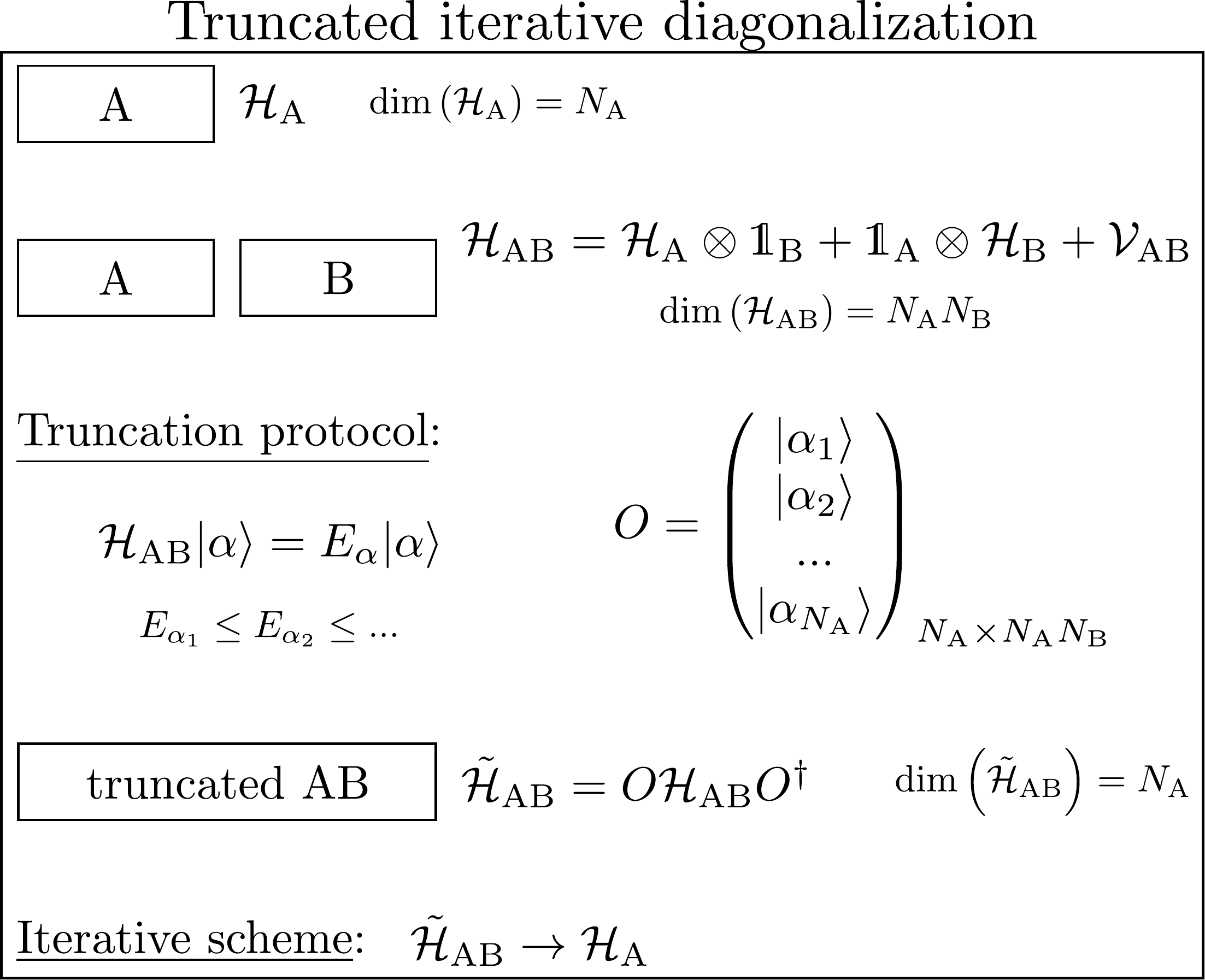}
 \caption{
 Schematic description of the truncated iterative diagonalization method.
 At every iteration, the system size is increased, whilst maintaining the dimension of the Hamiltonian matrix manageable for numerical diagonalization.
 This is achieved by projecting the basis of the enlarged system onto a truncated basis spanned by its lowest-energy eigenstates.
 The underlying assumption of this renormalization technique is that the low-energy states of the full system can be accurately described by the low-energy states of smaller blocks.
 }
 \label{fig:TID}
\end{figure}

The roots of DMRG can be traced back to a decimation procedure, to which we refer as TID.
Given a large, numerically intractable quantum system, the key idea of this approach is to divide it into smaller blocks that can be solved by exact diagonalization.
Combining these smaller blocks together, one at the time and integrating out the high-energy degrees of freedom, this renormalization technique arrives at a description of the full system in terms of a truncated Hamiltonian that can be diagonalized numerically.
The underlying assumption of this method is that the low-energy states of the full system can be accurately described by the low-energy states of smaller blocks.
The TID routine is one of the main steps in Wilson's numerical renormalization group~\cite{Wilson1975,Bulla2008}, which has had notable success in solving quantum impurity problems, such as the Kondo~\cite{Kondo1964} and the Anderson~\cite{Anderson1961} models.
As we shall point out below, TID was found to perform poorly for most quantum problems, working only for those where there is an intrinsic energy scale separation, such as quantum impurity models.

We now elaborate on the details of a TID implementation.
For that matter, let us consider TID as schematically described in Fig.~\ref{fig:TID}.
In the first step, we consider a small system A, with Hamiltonian $\mathcal{H}_\text{A}$, the dimension of which, $N_\text{A}$, is assumed to be manageable by numerical means.
In the next step, we increase the system size, forming what we denote by system AB, the Hamiltonian of which, $\mathcal{H}_\text{AB}$, has dimension $N_\text{A} N_\text{B}$ and is also assumed to be numerically tractable.
The Hamiltonian $\mathcal{H}_\text{AB}$ includes the Hamiltonians of the two individual blocks A and B, as well as their mutual interactions $\mathcal{V}_\text{AB}$.
Importantly, if we iterated the procedure at this step, it would be equivalent to doing exact diagonalization, in which case we would rapidly arrive at the situation where the dimension of the Hamiltonian matrix would increase to values that are too large to handle.
Instead, in the third step, we diagonalize $\mathcal{H}_\text{AB}$ and keep only its $N_\text{A}$ lowest-energy eigenstates~\footnote{More generally, the number of kept states does not need to be equal to $N_\text{A}$, but only small enough to stop the exponential growth and make the next iteration manageable.}.
These are used to form a rectangular matrix $O$, which can be employed to project the Hilbert space of the system AB onto a truncated basis spanned by its $N_\text{A}$ lowest-energy eigenstates, thereby integrating out the remaining higher-energy degrees of freedom.
As a consequence, it is possible to find an effective truncated version of any relevant operator defined in the system AB.
In particular, we can truncate $\mathcal{H}_\text{AB}$, obtaining an effective Hamiltonian $\tilde{\mathcal{H}}_\text{AB}$ with reduced dimension $N_\text{A}$, which can be used as the input for the first step of the next iteration.
This procedure is then iterated until the desired system size is reached.
As a final remark, we note that the matrices $O$ should be saved in memory at every iteration, as they are required to obtain the terms $\mathcal{V}_\text{AB}$, which we usually only know how to write in the original basis, as well as to compute expectation values of observables.

\begin{figure}
 \centering
 \includegraphics[width=0.6\columnwidth]{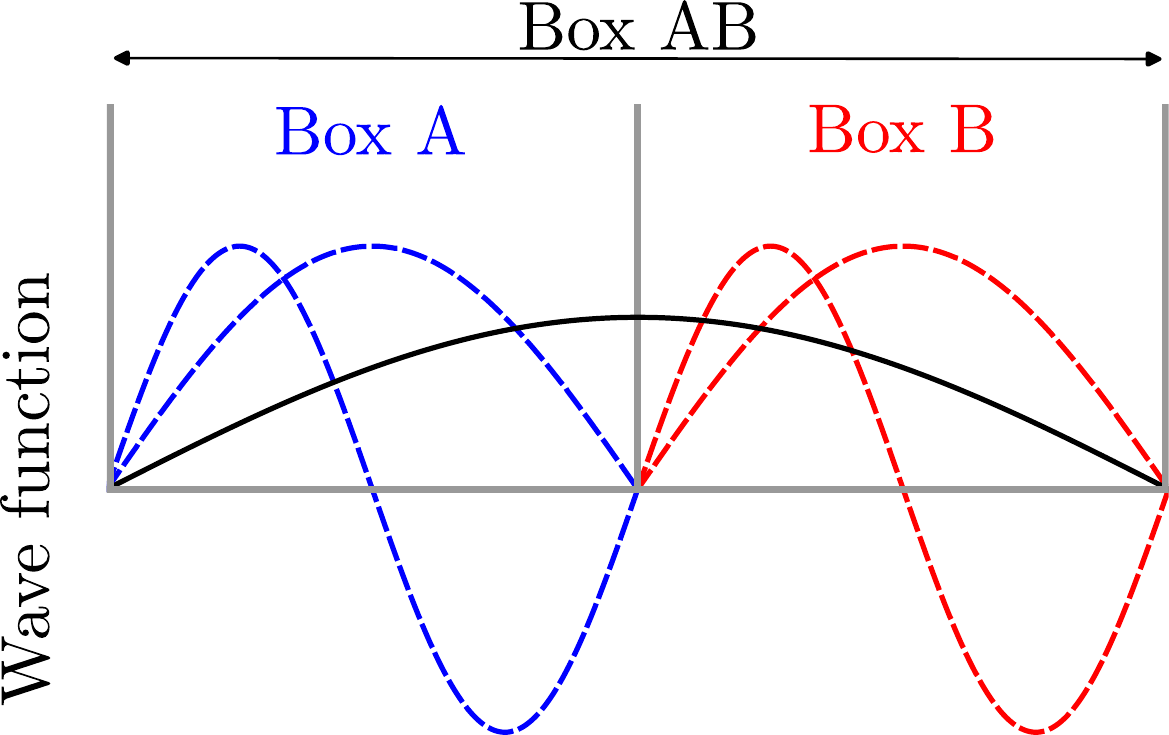}
 \caption{
 Illustration of the failure of the truncated iterative diagonalization method for the problem of a quantum particle in a box.
 The dashed blue (red) lines represent the two lowest-energy wave functions in the box A (B).
 The solid black line represents the lowest-energy wave function in the box AB.
 It is apparent that the lowest-energy state of the larger box cannot be obtained as a linear combination of a few low-energy states of the smaller boxes, thus leading to the breakdown of the principle that underpins the TID approach.
 }
 \label{fig:TID_fail_particle-in-box}
\end{figure}

Despite its rather intuitive formulation, TID turned out to yield poor results for most quantum many-body problems~\cite{White1993}.
In fact, White and Noack realized~\cite{White1992a} that this renormalization approach could not even be straightforwardly applied to one of the simplest (single-body) problems in quantum mechanics: the particle-in-a-box model (Fig.~\ref{fig:TID_fail_particle-in-box}).
Even though White and Noack managed to fix this issue by considering various combinations of boundary conditions, this observation was a clear drawback to the aspirations of TID, which motivated the search for a different method.
This culminated in the invention of DMRG, which is the focus of the next section.

\section{Original formulation of DMRG}\label{sec:DMRG_original}

\subsection{Infinite-system algorithm}
In 1992, Steven R. White realized that the eigenstates of the density matrix are more appropriate to describe a quantum system than the eigenstates of its Hamiltonian~\cite{White1992}.
This is the working principle of DMRG.
In this subsection, we consider the so-called infinite-system DMRG algorithm.
Even though it is possible to further improve this implementation scheme (see Section~\ref{sec:finite-DMRG}), it is an instructive starting point as it already contains the core ideas of DMRG.
Below, we introduce it in four steps.
First, we describe how to apply it, providing no motivation for its structure.
Second, we show, on the basis of the variational principle, that the truncation protocol prescribed by this method is optimal.
Third, we address its efficiency---i.e., how numerically affordable the truncation required for an accurate description of a large system is---, clarifying the models for which it is most suitable.
Fourth, we provide a pedagogical code implementation and discuss the results obtained.

\subsubsection{Description}\label{sec:iDMRG_description}

\begin{figure}
 \centering
 \includegraphics[width=\columnwidth]{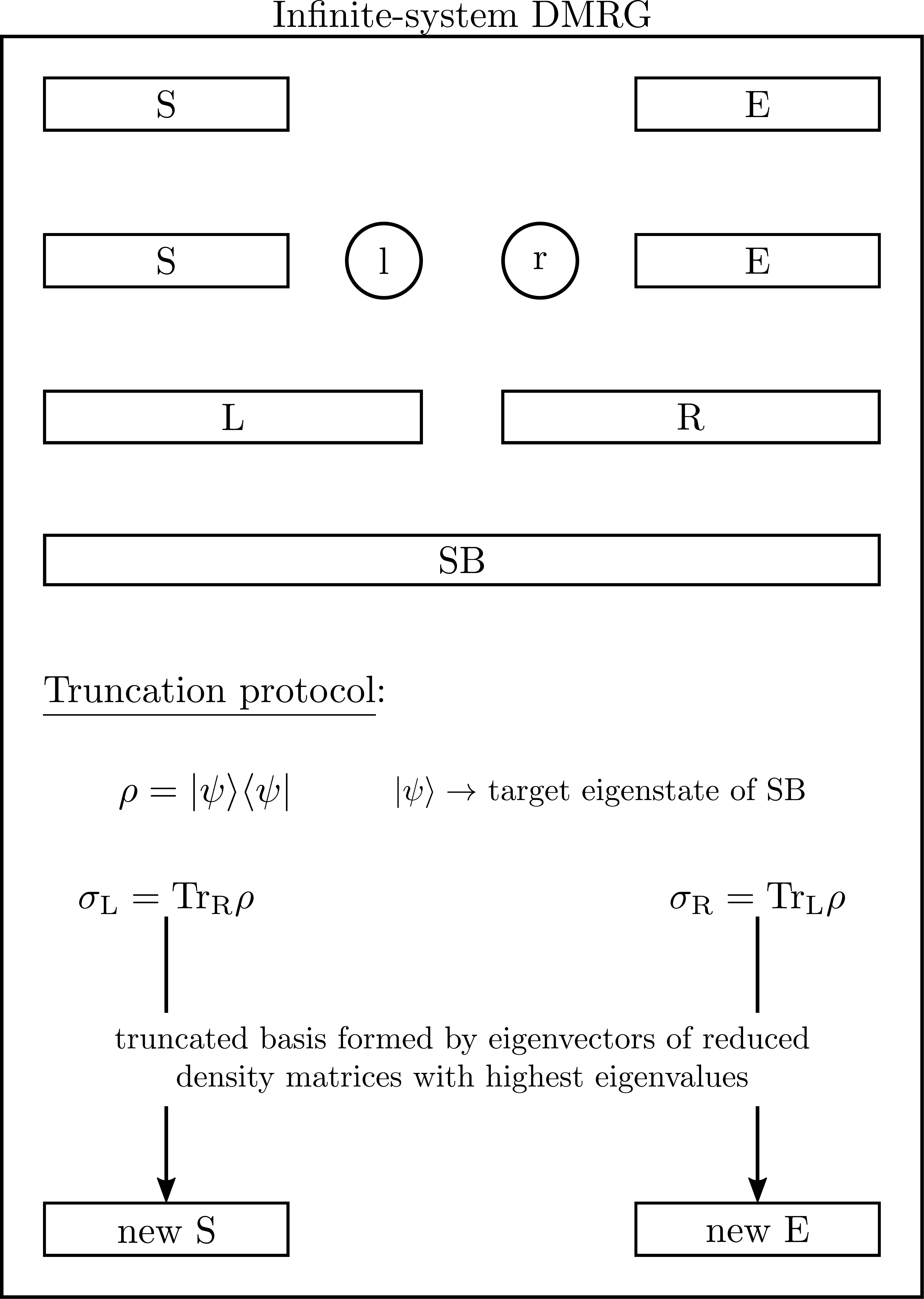}
 \caption{
 Schematic description of the infinite-system DMRG algorithm.
 Similarly to the TID approach, the system size is increased at every iteration while preventing an exponential growth of the dimension of its Hamiltonian matrix.
 The truncation employed involves the diagonalization of reduced density matrices, as their eigenvectors with highest eigenvalues are used to obtain an effective description of the enlarged system in a reduced basis.
 As shown in Section~\ref{sec:optimal_truncation_DMRG}, this truncation protocol is optimal and can, in principle, be applied to obtain the best approximation of any state $\vert \psi \rangle$ of an arbitrary quantum model.
In practice, however, this method is mostly useful to probe the low-energy states of 1D quantum problems with short-range interactions (see Section~\ref{subsubsec:efficiency}).
 }
 \label{fig:iDMRG}
\end{figure}

The infinite-system DMRG algorithm is schematically described in Fig.~\ref{fig:iDMRG}.
In the first step, we consider two blocks, denoted as S (system) and E (environment).
As we shall see, both blocks are part of the full system under study, so their designation is arbitrary.
Then, we increase the system size by adding two physical sites, one to each block, forming what we denote by blocks L (left) and R (right).
We proceed by building the block SB (superblock), which amounts to bundling the blocks L and R. 
The block SB is the representation of the full system that we intend to describe at every iteration.
It should be noted that all block aggregations imply that we account for the individual Hamiltonians of each block, plus their mutual interactions.
Finally, we move on to the truncations.
As a side remark, we point out that, if we truncated the blocks L and R using the corresponding low-energy states, forming new blocks S and E to use in the first step of the next iteration, this algorithm would be essentially equivalent to TID.
Alternatively, we diagonalize the block SB, and use one of its eigenstates $\vert \psi \rangle$ to build the density matrix $\rho = \vert \psi \rangle \langle \psi \vert$~\footnote{Here, we should choose the eigenstate $\vert \psi \rangle$ that we intend to obtain.
Most often, it will be the ground state or one of the lowest-energy eigenstates.
It is also possible to consider multiple target states $\vert \psi_n \rangle$, taking $\rho = \sum_n c_n \vert \psi_n \rangle \langle \psi_n \vert$, with $\sum_n c_n = 1$.
Drawbacks and best practices of this strategy are briefly discussed in Ref.~\cite{Schollwock2005}.}.
Then, we compute the reduced density matrices in the subspaces of the blocks L and R, $\sigma_\text{L/R} = \text{Tr}_\text{R/L} \rho$, diagonalize them, and keep their eigenvectors with highest eigenvalues.
These are used to truncate the blocks L and R, forming new blocks S and E that are taken as inputs of the first step in the next iteration.

\subsubsection{Argument for truncation}\label{sec:optimal_truncation_DMRG}
Here, we justify the truncation strategy prescribed above.
For that matter, let us consider an exact wave function of the block SB, written as
\begin{equation}
\vert \psi \rangle = \sum_{i_\text{L}=1}^{N_\text{L}} \sum_{i_\text{R}=1}^{N_\text{R}} \psi_{i_\text{L},i_\text{R}} \vert i_\text{L} \rangle \otimes \vert i_\text{R} \rangle,
\end{equation}
where $\vert i_\text{L} \rangle$ ($\vert i_\text{R} \rangle$) denotes a complete basis of the block L (R), with dimension $N_\text{L}$ ($N_\text{R}$).
We now propose a variational wave function of the form
\begin{equation}
\vert \tilde{\psi} \rangle = \sum_{\alpha_\text{L}=1}^{D_\text{L}} \sum_{i_\text{R}=1}^{N_\text{R}} c_{\alpha_\text{L},i_\text{R}} \vert \alpha_\text{L} \rangle \otimes \vert i_\text{R} \rangle,
\end{equation}
where $\vert \alpha_\text{L} \rangle$ denotes a truncated basis of the block L, with reduced dimension $D_\text{L} < N_\text{L}$.
The goal is to find the states $\vert \alpha_\text{L} \rangle$ and the variational coefficients $c_{\alpha_\text{L},i_\text{R}}$ that provide the best approximation of the truncated wave function $\vert \tilde{\psi} \rangle$ to the exact wave function $\vert \psi \rangle$, for a given $D_\text{L}$.
This can be achieved by minimizing $\| \vert \psi \rangle - \vert \tilde{\psi} \rangle \|^2$.

The exact wave function is normalized, i.e., $\langle \psi \vert \psi \rangle = 1$.
Using this property, we obtain
\begin{align}
& \| \vert \psi \rangle - \vert \tilde{\psi} \rangle \|^2 = 1 - \sum_{i_\text{L},i_\text{R},\alpha_\text{L}} \Big( \psi^*_{i_\text{L},i_\text{R}} c_{\alpha_\text{L},i_\text{R}} \langle i_\text{L} \vert \alpha_\text{L} \rangle \nonumber \\
& \quad + c^*_{\alpha_\text{L},i_\text{R}} \psi_{i_\text{L},i_\text{R}} \langle \alpha_\text{L} \vert i_\text{L} \rangle \Big) + \sum_{\alpha_\text{L},i_\text{R}} \vert c_{\alpha_\text{L},i_\text{R}} \vert^2, 
\label{eq:psi-psi'}
\end{align}
where we have also used the orthonormal properties of the basis states, e.g., $\langle i_\text{R} \vert i'_\text{R} \rangle = \delta_{i_\text{R},i'_\text{R}}$.
In order to minimize the previous expression, we impose that its derivative with respect to the variational coefficients $c_{\alpha_\text{L},i_\text{R}}$ (or $c^*_{\alpha_\text{L},i_\text{R}}$) must be zero.
This leads to
\begin{equation}
c_{\alpha_\text{L},i_\text{R}} = \sum_{i_\text{L}=1}^{N_\text{L}} \psi_{i_\text{L},i_\text{R}} \langle \alpha_\text{L} \vert i_\text{L} \rangle.
\label{eq:c_DMRG}
\end{equation}
Inserting Eq.~\eqref{eq:c_DMRG} into Eq.~\eqref{eq:psi-psi'}, we obtain
\begin{equation}
\| \vert \psi \rangle - \vert \tilde{\psi} \rangle \|^2 = 1 - \sum_{\alpha_\text{L}=1}^{D_\text{L}} \langle \alpha_\text{L} \vert \sigma_\text{L} \vert \alpha_\text{L} \rangle,
\label{eq:psi-psi'_trace}
\end{equation}
where we have introduced the reduced density matrix of the state $\vert \psi \rangle$ in the subspace of block L,
\begin{equation}
\sigma_\text{L} = \text{Tr}_\text{R} \rho = \sum_{i_\text{R}=1}^{N_\text{R}} \langle i_\text{R} \vert \rho \vert i_\text{R} \rangle,
\end{equation}
defined in terms of the full density matrix,
\begin{equation}
\rho = \vert \psi \rangle \langle \psi \vert.
\end{equation}

Looking at Eq.~\eqref{eq:psi-psi'_trace}, we observe that it involves a partial trace of the reduced density matrix $\sigma_\text{L}$ (note that $\sigma_\text{L}$ is an $N_\text{L} \times N_\text{L}$ matrix, but the sum over $\alpha_\text{L}$ runs over $D_\text{L}$ terms only).
Since $\sigma_\text{L}$ is a density matrix, its full trace must be equal to $1$, in which case the minimization of $\| \vert \psi \rangle - \vert \tilde{\psi} \rangle \|^2$ is accomplished by maximizing the partial trace of $\sigma_\text{L}$.
Per the Schur--Horn theorem~\cite{Schur1923,Horn1954}, the states $\vert \alpha_\text{L} \rangle$ that accomplish this are those that diagonalize $\sigma_\text{L}$ with highest eigenvalues $\lambda_{\alpha_\text{L}}$ (which are all non-negative, since any density matrix is positive semi-definite), i.e.,
\begin{equation}
\sigma_\text{L} \vert \alpha_\text{L} \rangle = \lambda_{\alpha_\text{L}} \vert \alpha_\text{L} \rangle, \quad \lambda_{1} \geq \lambda_{2} \geq ...,
\end{equation}
thus leading to
\begin{equation}
\| \vert \psi \rangle - \vert \tilde{\psi} \rangle \|^2 = 1 - \sum_{\alpha_\text{L}=1}^{D_\text{L}} \lambda_{\alpha_\text{L}}.
\label{eq:psi-psi'_final}
\end{equation}

Let us now put into words what we have just demonstrated.
Starting from an exact wave function $\vert \psi \rangle$, we can obtain a truncated (in the subspace of the block L) wave function $\vert \tilde{\psi} \rangle$ that best approximates $\vert \psi \rangle$ by going through the following protocol.
First, we build the density matrix $\rho = \vert \psi \rangle \langle \psi \vert$ and compute the reduced density matrix $\sigma_\text{L} = \text{Tr}_\text{R} \rho$.
Then, we diagonalize $\sigma_\text{L}$ and form a $D_\text{L} \times N_\text{L}$ matrix $O$ whose lines are the eigenvectors of $\sigma_\text{L}$ with highest eigenvalue.
Finally, $\vert \tilde{\psi} \rangle$ is obtained as $\vert \tilde{\psi} \rangle = O \vert \psi \rangle$.
Repeating the same strategy for the block R, for which the derivation is completely analogous, we arrive at the truncation scheme described in Section~\ref{sec:iDMRG_description}.

The calculation of $\| \vert \psi \rangle - \vert \tilde{\psi} \rangle \|^2$ at every iteration of the algorithm, using Eq.~\eqref{eq:psi-psi'_final}, can be used as a measure of the quality of the corresponding truncation.
Therefore, instead of fixing a given $D_\text{L}$, we can impose a maximum tolerance for $\| \vert \psi \rangle - \vert \tilde{\psi} \rangle \|^2$, 
obtaining an adaptive truncation scheme.
As a final remark, we note that, while the general derivation presented here applies to any state $\vert \psi \rangle$ of an arbitrary quantum problem, the efficiency of DMRG relies on how large $D_\text{L}$ must be to ensure that the truncation does not compromise the accurate quantitative description of the system under study. 
This subject is addressed below.

\subsubsection{Efficiency}\label{subsubsec:efficiency}
Recalling Eq.~\eqref{eq:psi-psi'_final}, it is apparent that the efficiency of DMRG relies on how fast the eigenvalues of the reduced density matrices decay for the quantum state $\ket{\psi}$ under study.
However, this property is, in general, unknown.
Instead, the entanglement entropy---for which general results are known or conjectured~\cite{Eisert2010}---can be used as a proxy, as explained below.

\begin{figure*}
 \centering
 \includegraphics[width=0.65\textwidth]{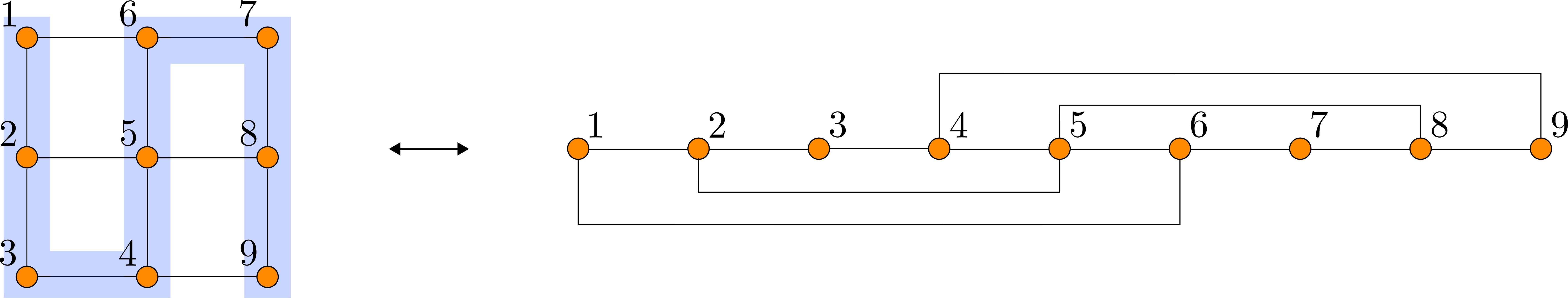}
 \caption{
 Relation between dimensionality and range of interactions on a lattice model.
 In the example depicted, a $3 \times 3$ two-dimensional square lattice with nearest-neighbor hopping terms is described as a 1D chain with hoppings up to fifth neighbors.
 In general, the same mapping applied to an $N \times N$ lattice leads to (nonlocal) hopping terms between sites separated by up to $2N-1$ units of the 1D chain.
 }
 \label{fig:2D_1D-long-range}
\end{figure*}

The blocks L and R form a bipartition of the full system, represented by the block SB.
We can therefore define the von Neumann entanglement entropy (of the state $\vert \psi \rangle$) between L and R as
\begin{align}
\mathcal{S} &\equiv \mathcal{S}(\sigma_\text{L}) = - \text{Tr} \left( \sigma_\text{L} \log_2 \sigma_\text{L} \right) \nonumber \\ 
&= \mathcal{S}(\sigma_\text{R}) = - \text{Tr} \left( \sigma_\text{R} \log_2 \sigma_\text{R} \right).
\end{align}
Focusing on the block L, without loss of generality, we write
\begin{equation}
\mathcal{S} = - \sum_{\alpha_\text{L}=1}^{N_\text{L}} \lambda_{\alpha_\text{L}} \log_2 \lambda_{\alpha_\text{L}} \simeq - \sum_{\alpha_\text{L}=1}^{D_\text{L}} \lambda_{\alpha_\text{L}} \log_2 \lambda_{\alpha_\text{L}},
\end{equation}
where we have restricted the sum over $\alpha_\text{L}$ to the $D_\text{L}$ highest eigenvalues of $\sigma_\text{L}$.
This approximation is valid since we are fixing $D_\text{L}$ so that $\| \vert \psi \rangle - \vert \tilde{\psi} \rangle \|^2 \simeq 0$, which implies, by virtue of Eq.~\eqref{eq:psi-psi'_final}, that the remaining eigenvalues are close to zero; given that $\lim\limits_{\lambda \to 0^+} \lambda \log_2 \lambda = 0$, it follows that the lowest eigenvalues of $\sigma_\text{L}$ can be safely discarded in the calculation of the entanglement entropy.
Within this assumption, it is also straightforward to check that $\mathcal{S}$ is maximal if $\lambda_{\alpha_\text{L}} = 1/D_\text{L}, \  \alpha_\text{L} = 1, 2, ..., D_\text{L}$, which allows us to write
\begin{equation}
\mathcal{S} \leq \log_2 D_\text{L},
\end{equation}
leading to
\begin{equation}
D_\text{L} \geq 2^\mathcal{S}.
\label{eq:bond_entropy_ineq}
\end{equation}
Using Eq.~\eqref{eq:bond_entropy_ineq}, we can make a rough estimate of the order of magnitude of $D_\text{L}$,
\begin{equation}
D_\text{L} \sim 2^\mathcal{S}.
\end{equation}

The scaling of $\mathcal{S}$ with the size of a translationally invariant quantum system is a property that is widely studied.
In particular, there are exceptional quantum states that obey the so-called area laws~\cite{Eisert2010}, meaning that $\mathcal{S}$, instead of being an extensive quantity~\footnote{Note that this situation, expected in the most general case, leads to an exponential scaling of $D_\text{L}$ with the system size, which is impractical for numerical purposes.}, is at most proportional to the boundary of the two partitions.
The area laws are commonly found to hold for the ground states of gapped Hamiltonians with local interactions~\cite{Eisert2010}; this result has been rigorously demonstrated in the 1D case~\cite{Hastings2007}.
It should also be noted that, for the ground states of 1D critical/gapless local models, the scenario is not dramatically worse as $\mathcal{S}$ is typically verified to scale only logarithmically with the chain length~\cite{Vidal2003,Latorre2004}.

In summary, considering the ground state of a local Hamiltonian describing a $\mathcal{D}$-dimensional system of size $\mathcal{L}$ in each dimension, we expect to have:
\begin{itemize}
    \item $\mathcal{S} \sim \text{const.}$, for 1D gapped systems. 
    This implies a favorable scaling $D_\text{L} \sim 2^\text{const.}$.
    
    \item $\mathcal{S} \sim c \log_2 \mathcal{L}$, for 1D gapless models. 
    This leads to $D_\text{L} \sim 2^{c \log_2 \mathcal{L}}$, yielding a power law in $\mathcal{L}$, which is usually numerically manageable in practical cases.

    \item $\mathcal{S} \sim \mathcal{L}^{\mathcal{D}-1}$, for gapped systems in $\mathcal{D}=2,3$ dimensions. 
    This implies $D_\text{L} \sim 2^{\mathcal{L}^{\mathcal{D}-1}}$, resulting in an exponential scaling that severely restricts the scalability of numerical calculations.
\end{itemize}
In short, we see that the truncation strategy employed in DMRG is in principle suitable for 1D quantum models (gapped or gapless), but not in higher dimensions.
Notable exceptions are two-dimensional problems whose solutions can be obtained or extrapolated from lattices where the size along one of the two dimensions is rather small, such as stripes or cylinders (see Ref.~\cite{Stoudenmire2012} for a review on the use of DMRG to study two-dimensional systems).
In fact, there is a relation between dimensionality and range of interactions in finite systems (Fig.~\ref{fig:2D_1D-long-range}), from which it also becomes apparent that DMRG is in practice only efficient when applied to models with short-range interactions.
Finally, it is reasonable to expect that the previous statements may hold not only for ground states but also for a few low-lying states.

\subsubsection{Code implementation}\label{subsubsec:original-DMRG_code}
In Supplementary Information, we present a didactic code implementation of the infinite-system DMRG algorithm, also made available at \url{https://github.com/GCatarina/DMRG_didactic}.
In this documented Jupyter notebook, written in Python, we focus on tackling spin-1 Heisenberg chains with open boundary conditions.
The generalization to different spin models is completely straightforward.
As for other types of quantum problems (e.g., fermionic models), this code can be readily used after simply defining the operators that appear in the corresponding Hamiltonian.
We also note that a slight modification of the algorithm has been proposed to better deal with periodic boundary conditions~\cite{Verstraete2004}.

\begin{figure}
 \centering
 \includegraphics[width=0.9\columnwidth]{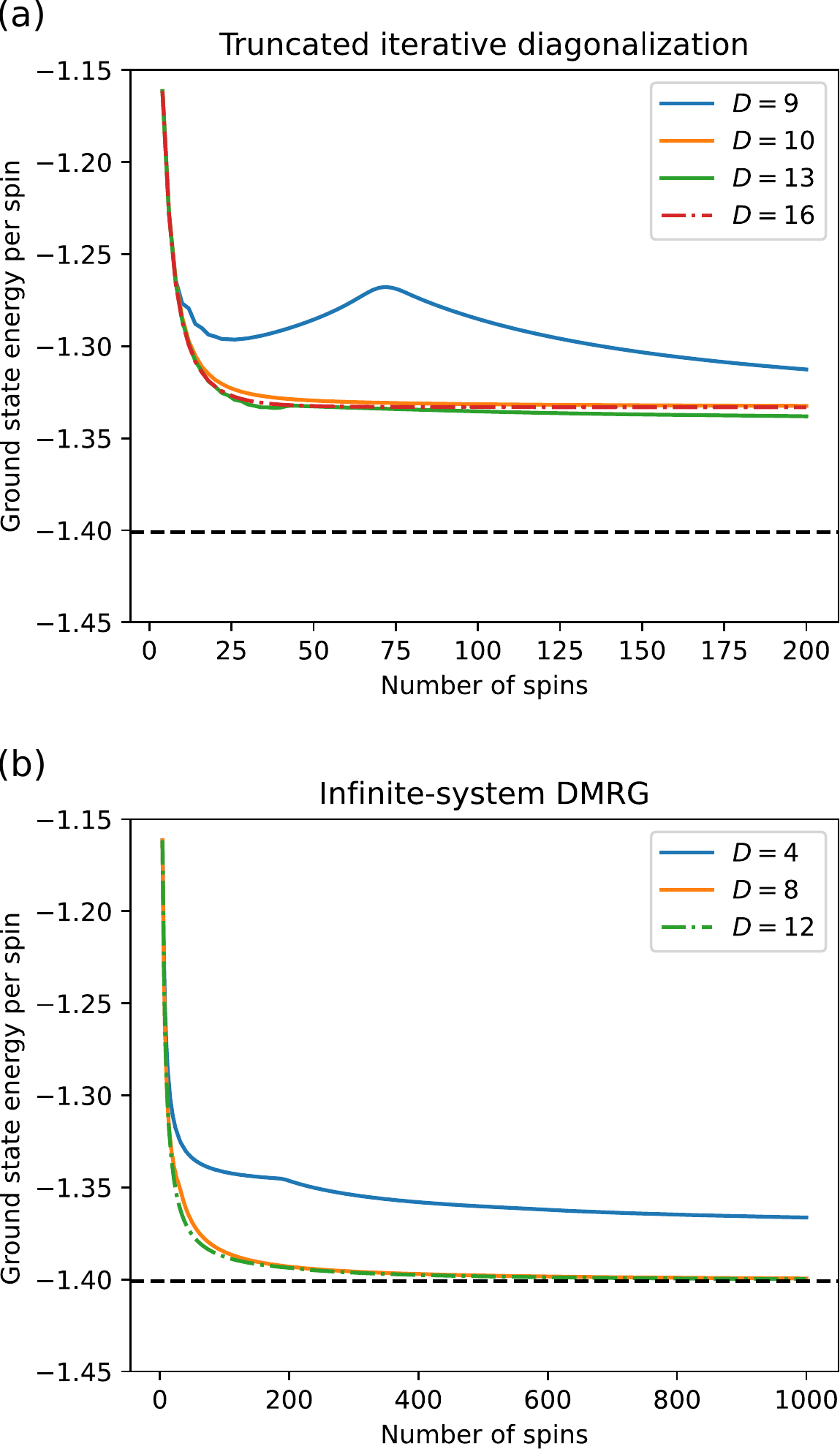}
 \caption{
 Benchmark results of truncated iterative diagonalization and infinite-system DMRG methods applied to open-ended spin-1 Heisenberg chains.
 Ground state energy per spin, as a function of the number of spins, obtained with TID (a) and infinite-system DMRG (b), for different values of $D$, which reflects the truncation employed, as described in the text.
 In both algorithms, every iteration implies the diagonalization of an Hamiltonian matrix of maximal dimension $9D^2 \times 9D^2$. 
 Larger matrices are allowed if degeneracies to within numerical precision are found at the truncation threshold, as explained in the code documentation.
 The dashed black line marks the known result in the thermodynamic limit~\cite{White1993a}.
 }
 \label{fig:TID-vs-iDMRG_Egs-Heis-spin1}
\end{figure}

\begin{figure}
 \centering
 \includegraphics[width=\columnwidth]{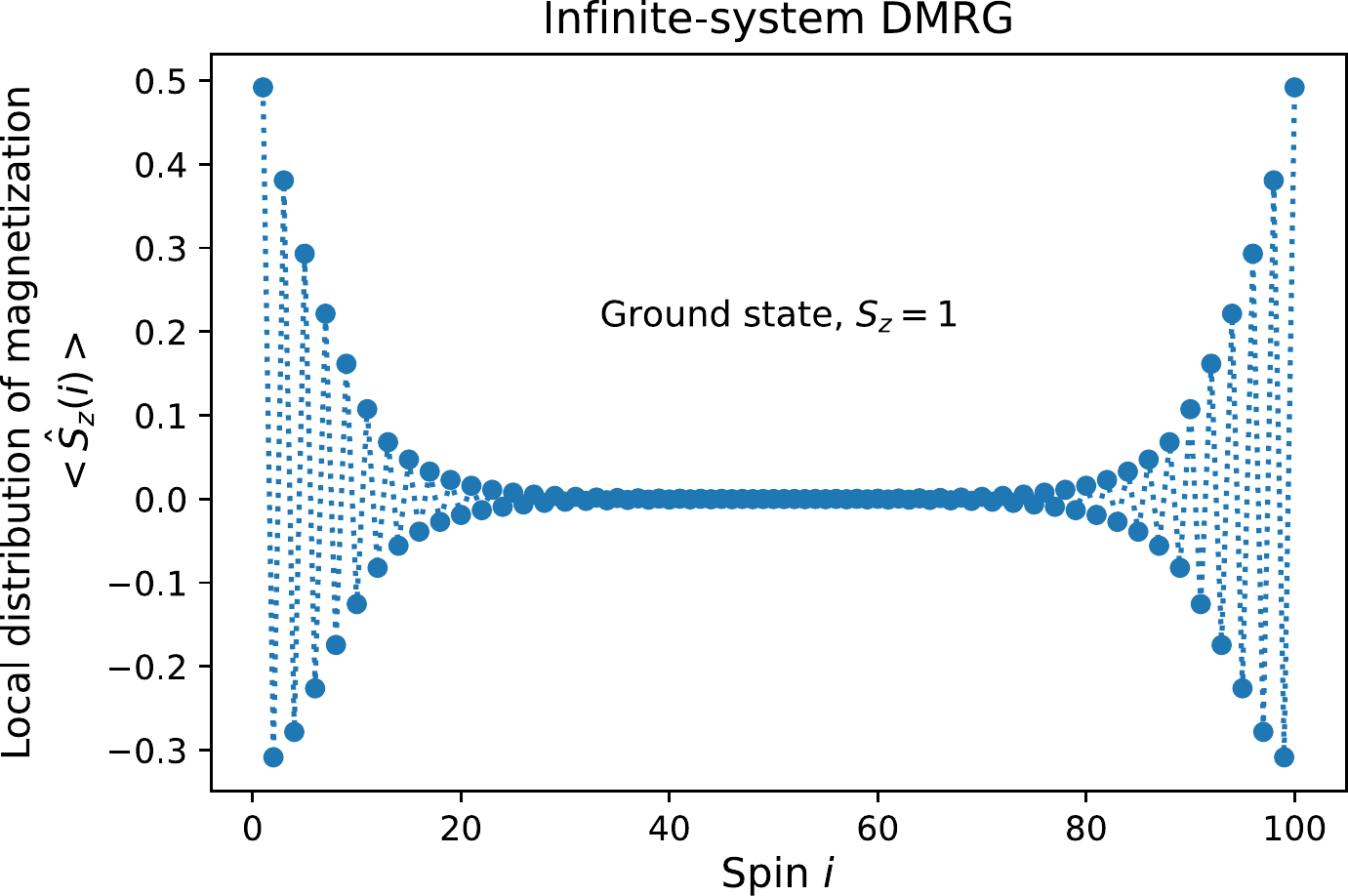}
 \caption{
 Magnetic properties of spin-1 Heisenberg chains computed by infinite-system DMRG.
 Local distribution of magnetization for the ground state with quantum number $S_z = +1$ (where $S_z$ denotes the total spin projection) of an open-ended chain composed of $100$ spins.
 The calculated local moments are exponentially localized at both edges of the chain, reflecting the fractionalization of the ground state into two effective spin-1/2 edge states.
 These results were obtained with an adaptive implementation in which the truncation error at every iteration was imposed to be below $10^{-4}$.
 A small Zeeman term was added to the Hamiltonian in order to target the $S_z = +1$ ground state.
 }
 \label{fig:iDMRG_Sz-fractional_Heis-spin1}
\end{figure}

For pedagogical purposes, our Jupyter notebook is structured in three parts.
First, we adopt the scheme described in Fig.~\ref{fig:iDMRG}, but make no truncations.
This is the same as doing exact diagonalization.
It is observed that, at every iteration, the running time of the code increases dramatically, reflecting the exponential wall problem.
Second, maintaining the same scheme, we make a truncation where the $D$ lowest-energy states of the block L (R) are used to obtain the new block S (E).
This is equivalent to the TID approach.
In Fig.~\ref{fig:TID-vs-iDMRG_Egs-Heis-spin1}(a), we plot the ground state energy per spin, as a function of the number of spins, obtained with this strategy, for different values of $D$.
Our calculations show a disagreement of at least $5 \%$ with the reference value~\cite{White1993a}, which does not appear to be overcome by considering larger values for $D$.
Therefore, we conclude that TID is not fully reliable for this problem.
Third, we implement the infinite-system DMRG, where we first set a fixed value for $D \equiv D_\text{L} = D_\text{R}$ in the truncations.
Computing the ground state energy per spin with this method, the results obtained are very close to the reference value, even for small values of $D$, as shown in Fig.~\ref{fig:TID-vs-iDMRG_Egs-Heis-spin1}(b).
For completeness, we also implement an adaptive version of the algorithm where the values of $D_\text{L/R}$ used at every iteration are set as to keep the truncation error, given by Eq.~\eqref{eq:psi-psi'_final}, below a certain threshold.
This adaptive implementation is used to compute the expectation values presented in Fig.~\ref{fig:iDMRG_Sz-fractional_Heis-spin1}, which show a known signature of the emergence of fractional spin-1/2 edge states in the model~\cite{White1993a}.

\subsection{Finite-system scheme}\label{sec:finite-DMRG}

\begin{figure*}
 \centering
 \includegraphics[width=0.75\textwidth]{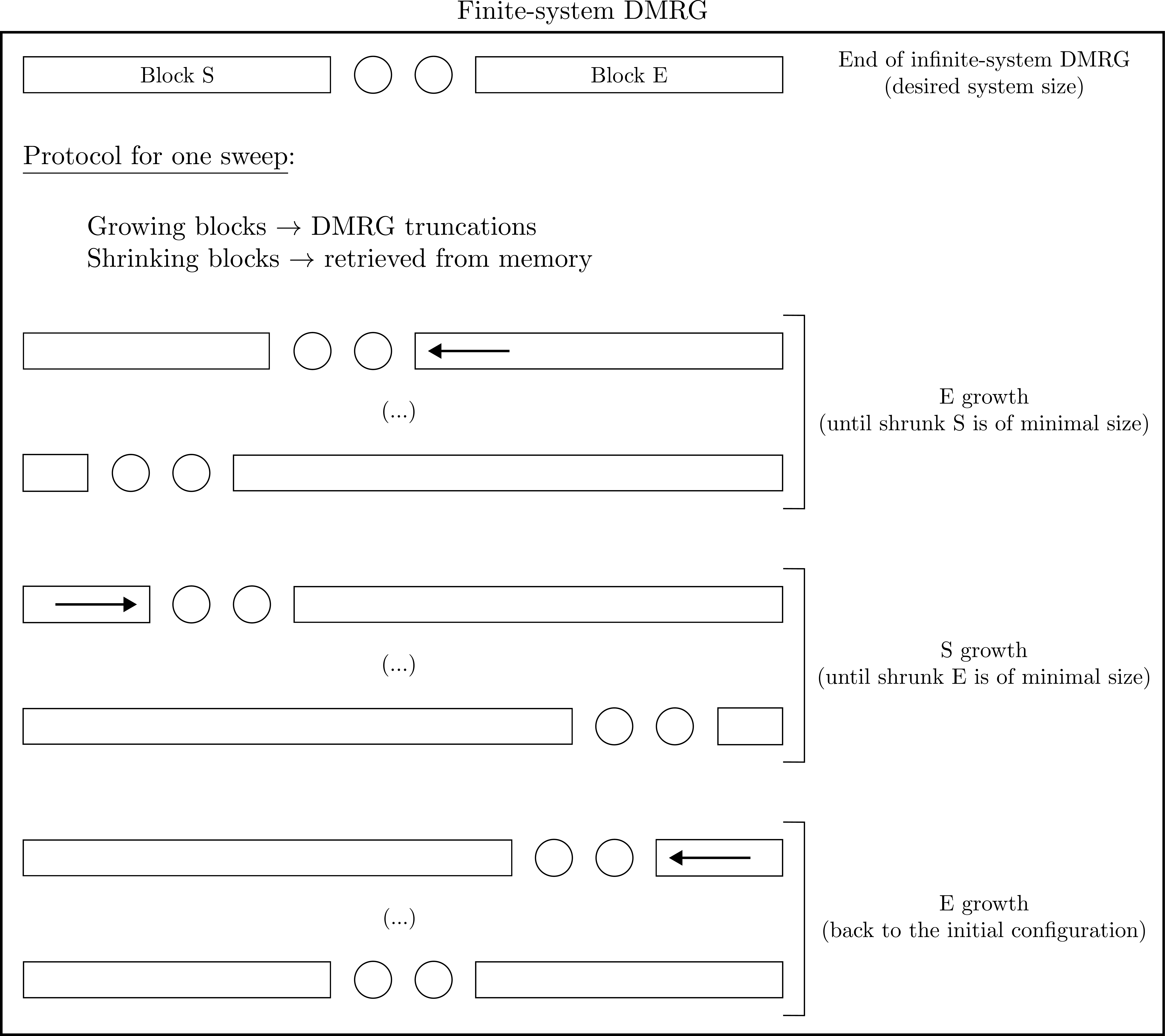}
 \caption{
 Breakdown of the finite-system DMRG routine.
 At the first stage, infinite-system DMRG is used to obtain an effective description for the target wave function of a system with desired size.
 This is followed by a sweeping protocol where one of the blocks is allowed to grow while the other is shrunk, thus keeping the total system size fixed.
 To prevent the exponential scaling, DMRG truncations (targeting the intended state) are employed for the growing blocks.
 The shrinking blocks are retrieved from memory, using stored data of the latest description of the block with such size (either from the infinite-system routine or from an earlier step of the sweeping procedure).
 The growth direction is reversed when the shrinking block reaches its minimal size.
 A typical strategy is to fix a maximal truncation error for the DMRG truncations, and perform sweeps until convergence in energy (and/or other physical quantities of interest) is attained; this approach ensures that the description of the target wave function is improved (or at least not worsened) at each step of the sweeping protocol.
 }
 \label{fig:finite-DMRG}
\end{figure*}

Within the infinite-system DMRG approach, the size of the system that we aim to describe increases at every iteration of the algorithm.
Therefore, the wave function targeted at each step is different.
This can lead to a poor convergence of the variational problem or even to incorrect results.
For instance, a metastable state can be favored by edge effects in the early DMRG steps, where the embedding with the environment is not so effective due to its small size, and the lack of ``thermalization'' in the following iterations may not allow for a proper convergence to the target state.

In this subsection, we present the so-called finite-system DMRG method, which manages to fix the aforementioned issues to a large extent.
The breakdown of this algorithm is shown in Fig.~\ref{fig:finite-DMRG}.
Its first step consists in applying the infinite-system routine to obtain an effective description for the target wave function of a system with desired size.
Then, a sweeping protocol is carried out to improve this description.
In this part, one of the blocks is allowed to grow, while simultaneously shrinking the other, thus keeping the overall system size fixed.
DMRG truncations (targeting the intended state) are employed for the growing blocks, whereas the shrinking blocks are retrieved from previous steps.
When the shrinking block reaches its minimal size, the growth direction is reversed.
A complete loop of this protocol, referred to as a \textit{sweep}, entails the shrinkage of the two blocks to their minimal sizes, and the return to the initial block configuration.
For a fixed truncation error, every step of a sweep must lead to a better (or at least equivalent) description of the target wave function; when the target is the ground state, this implies a variational optimization in which the estimated energy is a monotonically non-increasing function of the number of sweep steps performed.
This property is at the heart of the MPS formulation of DMRG (see Section~\ref{sec:5.1}).

As a final remark, we wish to clarify a few subtleties related to the variational character of DMRG.
For that matter, let us focus on the case where the target is the ground state wave function.
According to the derivation presented in Section~\ref{sec:optimal_truncation_DMRG}, it is straightforward to check that the DMRG truncations are variational in the number of kept states: a larger value of $D_\text{L/R}$ implies a better (or at least equivalent) description of the exact wave function, and hence a non-increasing energy estimation.
On top of that, we have just argued that, as long as we keep a fixed truncation error, the finite-system method is also variational in the number of sweeps.
Hence, the finite-system algorithm has an additional knob of optimization---the number of sweeps---that allows to improve the results of the infinite-system scheme.

\section{Tensor networks basics}\label{sec:TN_basics}

The modern formulation of DMRG is built upon tensor networks~\cite{Ostlund1995,Dukelsky1998,Schollwock2011}. 
Indeed, virtually all state-of-the-art implementations of DMRG~\cite{Hauschild2018,Fishman2022} make use of MPSs and MPOs. 
Although pedagogical reviews on these and other tensor networks are available~\cite{Orus2014,Bridgeman2017,Biamonte2017}, their scope goes far beyond DMRG, as they provide the reader with the required background to explore the broader literature on tensor network methods. 
Here, we take a more focused approach, giving the minimum necessary framework on tensor networks to understand the MPS-based version of the finite-system DMRG algorithm, which is discussed in detail in Section~\ref{sec:finite-DMRG_TN}.

\subsection{Diagrams and key operations}\label{subsec:tensorbasics}

\begin{figure*}
 \centering
 \includegraphics[width=0.7\linewidth]{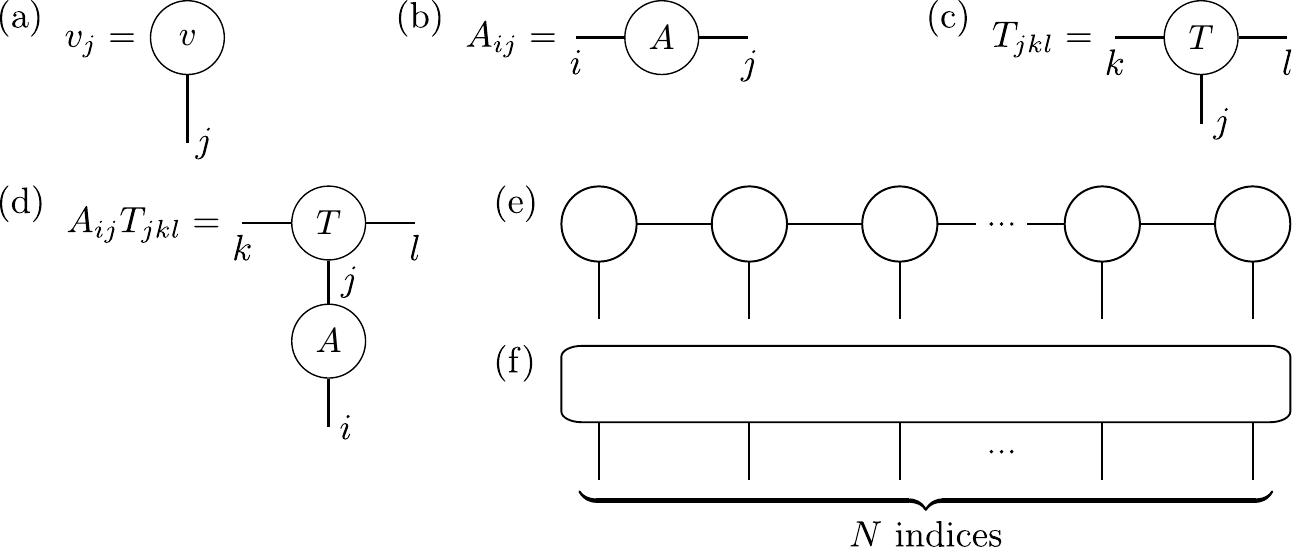}
 \caption{
 Diagrammatic representation of simple examples of tensors: (a) vector (i.e., rank-$1$ tensor), (b) matrix (i.e., rank-$2$ tensor), (c) rank-$3$ tensor. 
 Tensor networks are constructed by joining individual tensors, which is accomplished by contracting (i.e., summing over) indices in common. 
 (d) Example of contraction between rank-$2$ and rank-$3$ tensors.
 Common index $j$ is contracted. 
 Free indices $i$, $k$ and $l$ are represented through open legs. 
 (e) Example of canonical tensor network, MPS.
 Each local tensor has one free index.
 There is one contracted index (also known as bond) between every pair of adjacent tensors. 
 (f) Representation of generic rank-$N$ tensor.
 }
 \label{fig:tensors}
\end{figure*}

A tensor can be simply regarded as a mathematical object that stores information in a way that is determined by the number of indices $r \in \mathbb{N}^0$ (referred to as the \textit{rank} of the tensor), their dimensions $\{d_i\}_{i=1}^{r}$ (i.e., the $i^{\textrm{th}}$ index can take $d_i \in \mathbb{N}^+$ different values), and the order by which those indices are organized. 
The total number of entries of a tensor is $\prod_{i=1}^{r} d_i$. 
The most familiar examples of tensors are scalars (i.e., rank-$0$ tensors, each corresponding to a single number, thus not requiring any labels), vectors (i.e., rank-$1$ tensors, where every value is labelled by a single index that takes as many different values as the size of the vector), and matrices (i.e., rank-$2$ tensors, where every entry is characterized by two indices, one labelling the rows and another the columns). 
In general, each number stored in a rank-$r$ tensor is labelled in terms of an ordered array of $r$ indices, which can be regarded as its coordinates within the structure of the tensor. 
In Figs.~\ref{fig:tensors}(a)--(c), we show how tensors are represented diagrammatically. 

Although the number of indices, their dimensions, and the order by which they are organized are crucial to unambiguously label the entries of a tensor, these properties---to which we shall refer as the \textit{shape} of the tensor---are immaterial in the sense that we can fuse, split or permute its indices without actually changing the information contained within it. 
For clarity, let us consider the following $2 \times 4$ matrix $A_{\alpha \beta}$ with $\alpha \in \{0,1\}$, $\beta \in \{0,1,2,3\}$:
\vspace{-0.5cm}
\begin{figure}[H]
 \centering
 \includegraphics[width=0.5\linewidth]{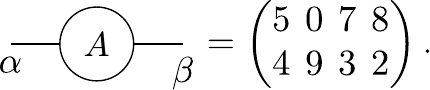}
\end{figure}
\vspace{-0.5cm}
\noindent We can \textit{reshape} this rank-$2$ tensor by \textit{fusing} its two indices, yielding the $8$-dimensional vector $A_{(\alpha, \beta)} \equiv A_{\gamma}$ if the row index $\alpha$ is chosen to precede the column index $\beta$, 
\vspace{-0.5cm}
\begin{figure}[H]
 \centering
 \includegraphics[width=0.75\linewidth]{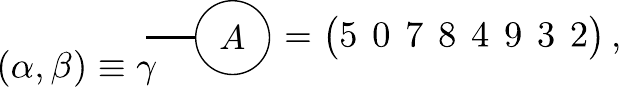}
\end{figure}
\vspace{-0.5cm}
\noindent or $A_{(\beta, \alpha)} \equiv A_{\delta}$ if $\beta$ takes precedence,
\vspace{-0.5cm}
\begin{figure}[H]
 \centering
 \includegraphics[width=0.75\linewidth]{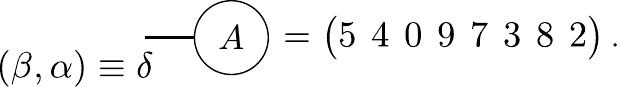}
\end{figure}
\vspace{-0.5cm}
\noindent Likewise, we can \textit{split} the $d=4$ column index $\beta$ into two $d=2$ indices, $\beta_1, \beta_2 \in \{0,1\}$, corresponding to the least and most significant digits of the binary decomposition of $\beta$, respectively. 
This yields the rank-$3$ tensor
\vspace{-0.5cm}
\begin{figure}[H]
 \centering
 \includegraphics[width=0.9\linewidth]{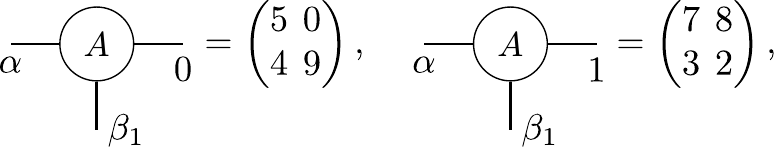}
\end{figure}
\vspace{-0.5cm}
\noindent where $\beta_2$ corresponds to the rightmost leg in the diagrams above. 
Alternatively, we can leave the rank unchanged, \textit{permuting} the row and column indices to yield the $4 \times 2$ transpose matrix
\vspace{-0.5cm}
\begin{figure}[H]
 \centering
 \includegraphics[width=0.45\linewidth]{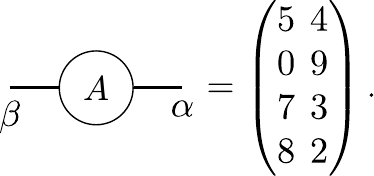}
\end{figure}
\vspace{-0.5cm}
\noindent In all three cases, even though we end up with tensors of different shape, all of them store exactly the same content as the original matrix, albeit in a different way. 
This is the key point: reshaping a tensor (by fusing or splitting indices) or simply permuting its indices merely restructures how the information is stored, leaving the information itself unaffected.
In the context of numerical implementations, we note that these tensor operations can be applied to arbitrary-rank tensors via standard built-in functions (e.g., \texttt{numpy.reshape} and \texttt{numpy.transpose} in Python). 
The time complexity of reshaping a tensor or permuting its indices is essentially negligible, as these operations just modify a flag associated with the tensor that defines its shape rather than actually moving its elements around in memory.

Thus far, we have only considered isolated tensors. 
However, based on the diagrammatic representations illustrated in Figs.~\ref{fig:tensors}(a)--(c), where each index corresponds to a leg, we can think of joining two individual tensors by linking a pair of legs, one from each tensor, as shown in Fig.~\ref{fig:tensors}(d). 
Algebraically, such link/bond corresponds to a sum over a common index shared by the two tensors; the outcome of this operation can be explicitly obtained in Python via \texttt{numpy.einsum}.
Of course, this process can be generalized to an arbitrary number of tensors, resulting in \textit{tensor networks} of arbitrary shapes and sizes. 
Here, we will focus on the so-called \textit{matrix product states} (MPSs), relevant for DMRG.
A diagram of an MPS is shown in Fig.~\ref{fig:tensors}(e); it comprises both free indices (i.e., open legs) and contracted indices (i.e., bonds). 
The elements of an MPS are uniquely identified by the free indices, but, unlike the case of an isolated tensor, their values are not immediately available, as the contracted indices must be summed over to obtain them.
In the context of DMRG, an MPS with $N$ free/physical indices is typically used to represent a quantum state of a system with $N$ sites.

\begin{figure*}
 \centering
 \includegraphics[width=0.9\linewidth]{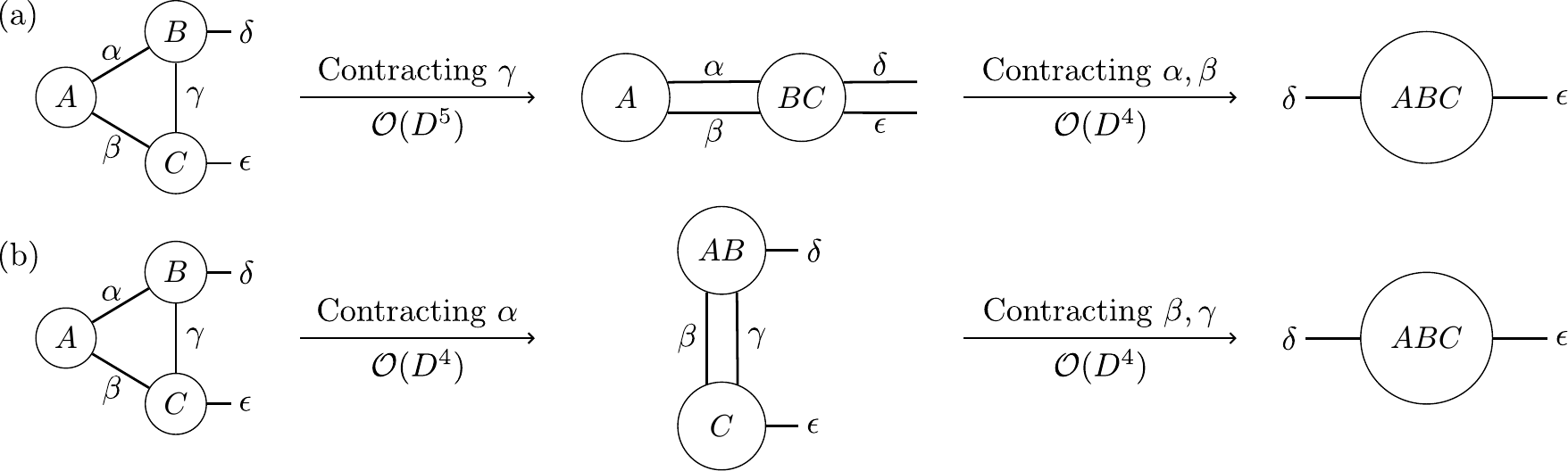}
 \caption{
 Comparison of two strategies to contract a tensor network comprising three tensors. 
 All indices, both free and contracted, are assumed to have dimension $D$ for the purpose of estimating scaling of cost of contractions. 
 (a) First, index $\gamma$ linking tensors $B$ and $C$ is contracted, and then indices $\alpha$ and $\beta$ are summed over, yielding an overall cost of $\mathcal{O}(D^5)$. 
 (b) First, index $\alpha$ linking tensors $A$ and $B$ is contracted, and then indices $\beta$ and $\gamma$ are summed over, resulting in $\mathcal{O}(D^4)$ cost. 
 Even though both strategies yield the same outcome, (b) is preferred, since its execution time scales more favorably with the index dimension $D$.
 }
 \label{fig:contractcost}
\end{figure*}

Even though the order by which sums over contracted indices are performed does not affect the obtained result, different orders may produce substantially different times of execution, especially if the tensor networks in question are large. 
For the 1D tensor networks herein considered, the type of contractions that we need to deal with are essentially those shown in Fig.~\ref{fig:contractcost}, for which there are two possible contraction strategies.
Contracting multiple bonds of a tensor network essentially amounts to performing nested loops. 
When we sum over a given contracted index, corresponding to the current innermost loop, we effectively have to fix the dummy variables of the outer loops. 
However, all possible values that such dummy variables can take must be considered. 
In the scheme of Fig.~\ref{fig:contractcost}(a), we first contract the $D$-dimensional bond linking tensors $B$ and $C$, which involves order $\mathcal{O}(D)$ operations on its own, but we must repeat this for all possible combinations of values of all other indices of tensors $B$ and $C$, which are $\mathcal{O}(D^4)$, yielding a total scaling of $\mathcal{O}(D^5)$. 
The second step contracts both bonds linking $A$ to $BC$, taking $\mathcal{O}(D^4)$ operations. 
For Fig.~\ref{fig:contractcost}(b), in turn, contracting first the bond between $A$ and $B$ takes $\mathcal{O}(D^4)$ operations, and the same scaling is obtained for the second step. 
Hence, (b) has an overall cost of $\mathcal{O}(D^4)$, which is more favorable than the $\mathcal{O}(D^5)$ scaling of (a). 
In general, the problem of determining the optimal contraction scheme is known to be NP-hard~\cite{ChiChung1997,Pfeifer2014}, but this issue only arises in two and higher dimensions. 
For our purposes, the cases described above are all we need to know about tensor network contractions.

Tensor networks can be regarded as tensors with internal structure. 
Therein lies their great virtue: such internal structure allows for a compact storage of information, which greatly reduces the memory requirements of the variational methods that use these tensor networks as their ans\"{a}tze. 
For concreteness, let us compare the $N$-site MPS shown in Fig.~\ref{fig:tensors}(e) to an isolated rank-$N$ tensor (resultant, e.g., from contracting all the bonds of the $N$-site MPS), shown in Fig.~\ref{fig:tensors}(f). 
Assuming free and contracted indices have dimension $d$ and $D$, respectively, while the isolated tensor requires storing a total of $d^{N}$ numbers in memory, the MPS only involves saving the entries of $N-2$ rank-$3$ $D \times d \times D$ tensors in the bulk and $2$ rank-$2$ $d \times D$ tensors at the ends, yielding $\mathcal{O}(N D^2 d)$ numbers saved in memory. 
In other words, the memory requirements of methods based on MPSs scale linearly with the system size $N$, in contrast with the exponential scaling associated with an unstructured tensor.

\subsection{Singular value decomposition}\label{subsec:svd}

The success of the original formulation of DMRG in tackling quantum many-body problems in a scalable way rests upon the projection of the Hilbert space onto the subspace spanned by the highest-weight eigenstates of the reduced density matrix on either side of the bipartition considered. 
In the MPS-based formulation, the analogue operation (see Section \ref{sec:5.2}) corresponds to the \textit{singular value decomposition} (SVD) of the local tensors that compose the MPS.

SVD consists of factorizing any $m \times n$ real or complex matrix $M$ in the form $M = \mathcal{U} \mathcal{S} \mathcal{V}^{\dagger}$, where $\mathcal{U}$ and $\mathcal{V}$ are $m \times m$ and $n \times n$ unitary matrices, respectively, and $\mathcal{S}$ is an $m \times n$ matrix with non-negative real numbers (some of which possibly zero) along the diagonal and all remaining entries equal to zero:  
\vspace{-0.5cm}
\begin{figure}[H]
 \centering
 \includegraphics[width=\linewidth]{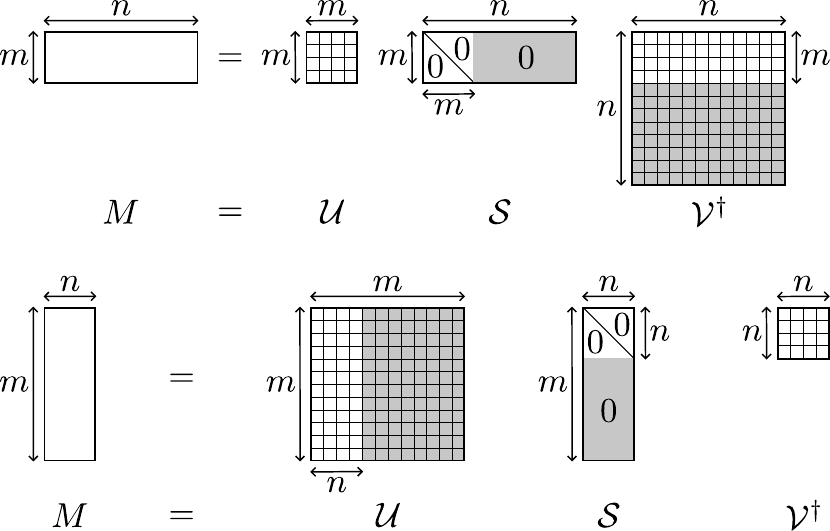}
\end{figure}
\vspace{-0.5cm}
\noindent In the schematic representations of SVD above, the parallel horizontal and vertical lines forming the grids within $\mathcal{U}$ and $\mathcal{V}^{\dagger}$ serve to illustrate that the respective rows and columns form an orthonormal set, which is the defining property of a unitary matrix. 
As highlighted by the shaded regions, all entries of the the last $n - m$ columns (if $m < n$) or the last $m - n$ rows (if $m > n$) of $\mathcal{S}$ are zero, so we can remove such redundant information by truncating $\mathcal{U}$, $\mathcal{S}$ and $\mathcal{V}^{\dagger}$ (the truncated versions of which we write as $U$, $S$ and $V^\dagger$) accordingly: 
\vspace{-0.5cm}
\begin{figure}[H]
 \centering
 \includegraphics[width=\linewidth]{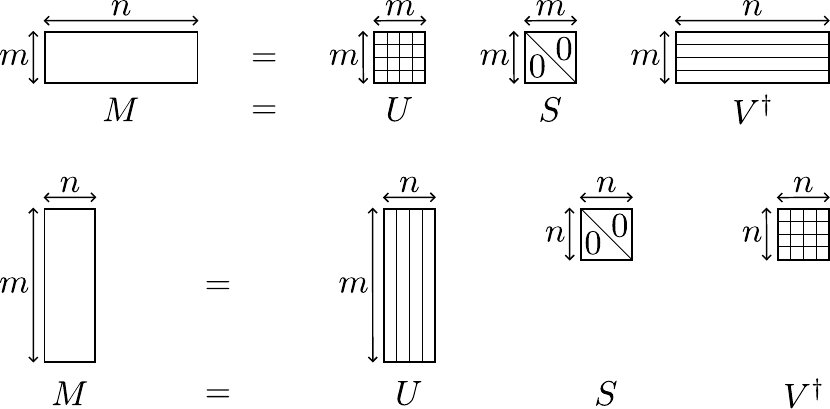}
\end{figure}
\vspace{-0.5cm}
\noindent This is the so-called \textit{thin} or \textit{reduced} SVD, as opposed to the \textit{full} SVD described earlier. 
Both are implemented in Python via \texttt{numpy.linalg.svd}, setting the Boolean input parameter \texttt{full\_matrices} appropriately. 
Henceforth, unless stated otherwise, we shall consider the thin SVD, as it yields the most compact factorization of the original matrix $M$.

A brief overview of some terminology related to SVD is in order. 
First, in the thin SVD diagrams above, $V^{\dagger}$ in the $m < n$ case and $U$ in the $m > n$ case are rectangular matrices, and therefore neither is unitary. 
Nevertheless, as illustrated through the parallel lines, the rows of $V^{\dagger}$ in the $m < n$ case and the columns of $U$ in the $m > n$ case still form an orthonormal basis, so the former is said to be \textit{right-normalized} (i.e., $V^{\dagger} (V^{\dagger})^{\dagger} = V^{\dagger} V = \mathds{1}$) and the latter \textit{left-normalized} (i.e., $U^{\dagger} U = \mathds{1}$). 
The columns of $U$ and the rows of $V^{\dagger}$ are referred to as left- and right-singular vectors. 
The diagonal entries of the $\textrm{min}\{ m, n\} \times \textrm{min}\{ m, n\}$ matrix $S$ are called \textit{singular values}. 
The Schmidt rank $r_{\text{S}} \leq \textrm{min}\{ m, n\}$ is the number of nonzero singular values. 
By exploiting the gauge freedom of SVD (see Appendix~\ref{gauge_SVD}), the singular values are conventionally stored in descending order, which is useful when truncations are considered, as explained below.

\begin{figure*}
 \centering
 \includegraphics[width=\textwidth]{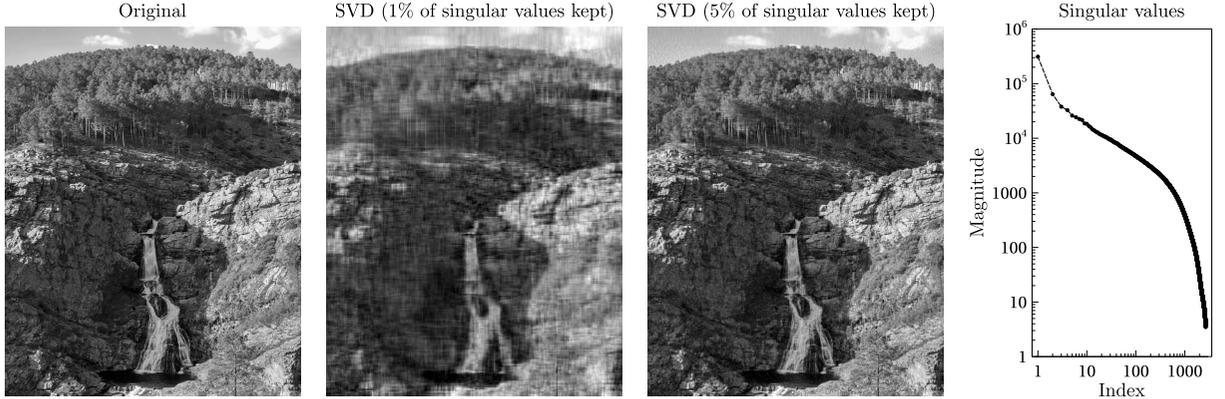}
 \caption{
 Singular value decomposition for image compression.
 The original photo (taken at Fisgas de Ermelo, Portugal) is stored as a $3335 \times 2668$ matrix, where each entry corresponds to a pixel and the values encode the grayscale color.
 The compressed images are obtained by applying SVD to this matrix, keeping only the highest singular values (namely, $1\%$ and $5\%$ of the total $2668$ singular values).
 The distribution of the singular values is shown in the rightmost panel.
 }
 \label{fig:SVD_image-compression}
\end{figure*}

The application of the thin SVD to a rectangular matrix allows for a trivial truncation of the bond dimensions between the factorized matrices.
Further truncations can be implemented by discarding singular values of negligible magnitude.
If the discarded singular values are zero, this procedure is exact.
Otherwise, some information is lost, but the strategy of discarding the lowest singular values is known to yield the optimal truncation~\cite{Eckart1936,Mirsky1960}.
Therefore, SVD is widely used for data compression, being particularly efficient in cases where the singular values decay rapidly.
In Fig.~\ref{fig:SVD_image-compression}, we show such an example, where SVD is used to compress a black-and-white photograph.
We observe that, by keeping only the $1\%$ highest singular values, the image obtained already exhibits most of the features of the original photo, though noticeably blurred out. 
This blur is significantly reduced when the number of kept singular values is increased to just $5\%$.

\begin{figure*}
 \centering
 \includegraphics[width=\linewidth]{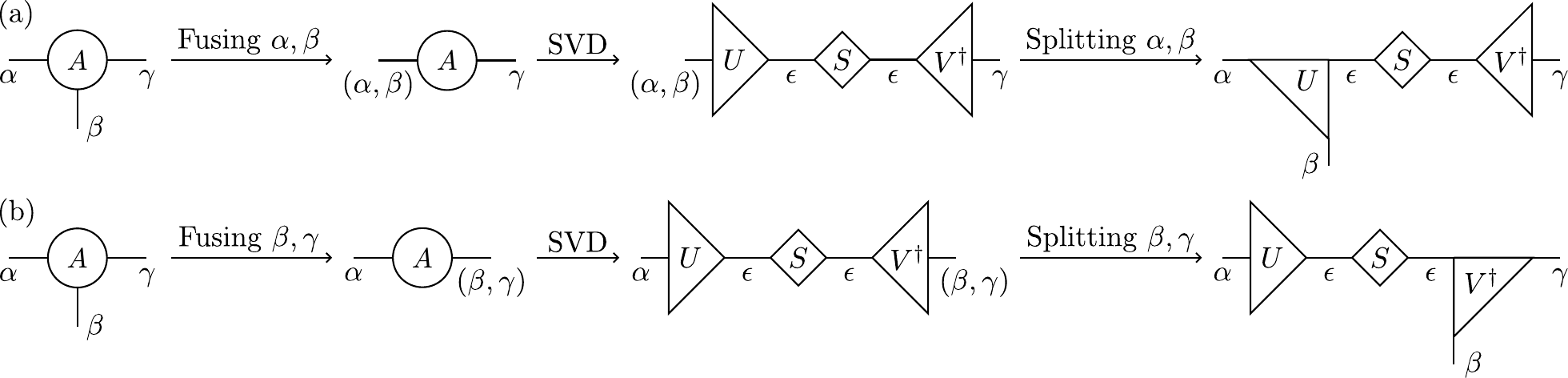}
 \caption{
 Singular value decomposition of rank-$3$ tensor $A$ belonging to a matrix product state. 
 (a) Central/physical index $\beta$ is fused with leftmost index $\alpha$ to yield left-normalized tensor $U$ at current site after SVD and index splitting. 
 The remaining $S V^{\dagger}$ is contracted with the local tensor that appears to the right of $A$ in the MPS.
 (b) Central/physical index $\beta$ is fused with rightmost index $\gamma$ to yield right-normalized tensor $V^{\dagger}$ at current site after SVD and index splitting. 
 The remaining $U S$ is contracted with the local tensor that appears to the left of $A$ in the MPS.
 Triangular shapes indicate left- and right-normalization of $U$ and $V^\dagger$, respectively.
 Diamond-shaped diagram illustrates that $S$ is diagonal.
 } 
 \label{fig:SVD_rank3}
\end{figure*}

A wave function can always be exactly represented by an MPS, although this will generally entail an exponential growth of the bond dimensions from the ends towards the center of the MPS (see Section~\ref{subsubsec:MPS_general-wavefunction}).
Within the context of MPS-based DMRG, SVD is adopted both to truncate the bond dimensions of the MPS and to transform it into convenient canonical forms, which we shall introduce in Section~\ref{subsubsec:canonical_forms}. 
However, since SVD is a linear algebraic method, it applies to matrices and not to the rank-$3$ tensors found in the non-terminal sites of an MPS. 
As a result, these tensors have to be reshaped by fusing two indices. 
There are two possibilities for this, depending on which leg we choose to fuse the physical index with (Fig.~\ref{fig:SVD_rank3}). 
In Fig.~\ref{fig:SVD_rank3}(a), we end up with a left-normalized tensor $U$ at the current site, with the remaining $S V^{\dagger}$ being contracted with the next local tensor to the right of the MPS. 
In Fig.~\ref{fig:SVD_rank3}(b), the right-normalized tensor $V^{\dagger}$ is the final form of the tensor at the current site, and $U S$ is contracted with the next local tensor to the left. 
The expressive power of an MPS is determined by the bond dimension cutoff $D$, which sets the maximum size of the contracted indices (e.g., $\alpha$, $\gamma$, and $\epsilon$ in Fig.~\ref{fig:SVD_rank3}). 
The dimension $d$ of the physical indices (e.g., $\beta$ in Fig.~\ref{fig:SVD_rank3}) is fixed by the local degrees of freedom of the problem under consideration (e.g., $d = 2s+1$ for a spin-$s$ quantum model). 
As a result, in Figs.~\ref{fig:SVD_rank3}(a)--(b), both dimensions of the matrix resulting from reshaping the rank-$3$ tensor are $\mathcal{O}(D)$. 
Computing the SVD of an $m \times n$ matrix (with $m > n$) takes $\mathcal{O}(m^2 n + n^3)$ floating-point operations~\cite{Golub1996}.
Hence, within the context of MPS-based DMRG, the time complexity of SVD is $\mathcal{O}(D^3)$.

\subsection{Matrix product states}

This subsection introduces the key operations required to manipulate MPSs. 
In particular, we discuss how to compute overlaps between two MPSs and expectation values of MPSs for local operators~\footnote{A more general discussion of the computation of expectation values with MPSs is deferred to the next subsection, where we introduce MPOs.}. 
Three MPS canonical forms that simplify some of these computations are introduced; the construction of all of them merely involves a sequential sitewise application of SVD, as described in Fig.~\ref{fig:SVD_rank3}.
For completeness, we also explain how to obtain an MPS representation of a general wave function, even though this procedure is not essential for DMRG.
In general, we shall consider $N$-site MPSs with bond dimension $D$ and physical index dimension $d$.

\subsubsection{Overlaps}\label{subsec:MPSoverlaps}

\begin{figure}
 \centering
 \includegraphics[width=0.9\linewidth]{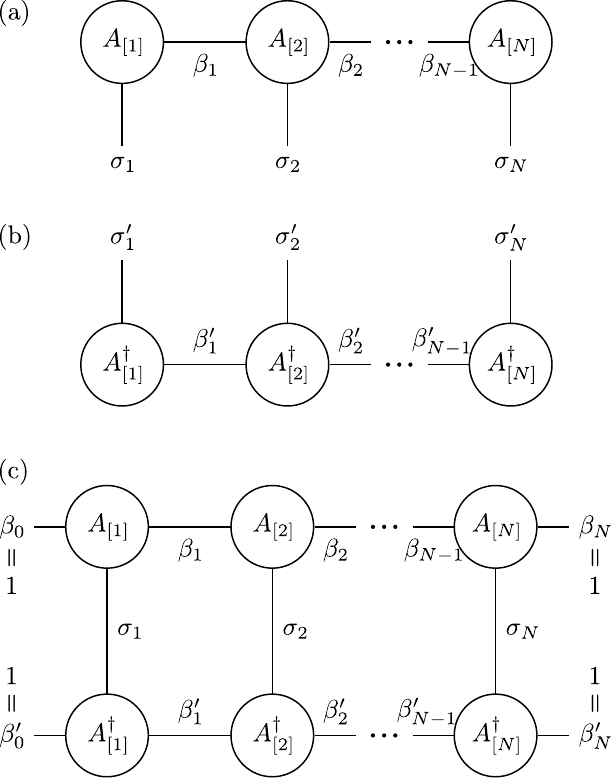}
 \caption{
 Diagrammatic representation of (a) MPS for ket $\ket{\psi}$, (b) MPS for bra $\bra{\psi}$, and (c) contraction of two previous MPSs to compute norm $\braket{\psi}$. 
 In (c), singleton dummy indices $\beta_0$, $\beta'_0$, $\beta_N$ and $\beta'_N$ were added on either side of both MPSs to ease discussion of efficient method to contract tensors down to scalar $\braket{\psi}$ (see Fig.~\ref{fig:zippermain}).
 }\label{fig:MPSoverlap}
\end{figure}

Using Dirac's bra-ket notation, the MPS representations of a ket $\ket{\psi}$ and its bra $\bra{\psi}$ are shown in Figs.~\ref{fig:MPSoverlap}(a)--(b), respectively. 
The diagrammatic representation of the norm of this state, $\braket{\psi}$, amounts to linking the two MPSs by joining the physical indices $\{ \sigma_i \}_{i = 1}^{N}$, as shown in Fig.~\ref{fig:MPSoverlap}(c). 
The question, then, is how to contract such tensor network to arrive at the scalar $\braket{\psi}$. 
A na\"{i}f approach would be to fix the same set of physical indices in the bra and the ket ($\sigma_i = \sigma'_i$), contract the remaining bonds ($N-1$ at the ket and $N-1$ at the bra), multiply the scalars obtained in the bra and the ket, and then sum over all possible values of the physical indices. 
The problem, however, is that $\{ \sigma_i \}_{i = 1}^{N}$ take $d^{N}$ different values, so this would be exponentially costly in $N$. 
Fortunately, there is a contraction scheme linear in $N$ that resembles the process of \textit{closing a zipper}~\cite{vonDelft2020}.

\begin{figure}
 \centering
 \includegraphics[width=0.95\linewidth]{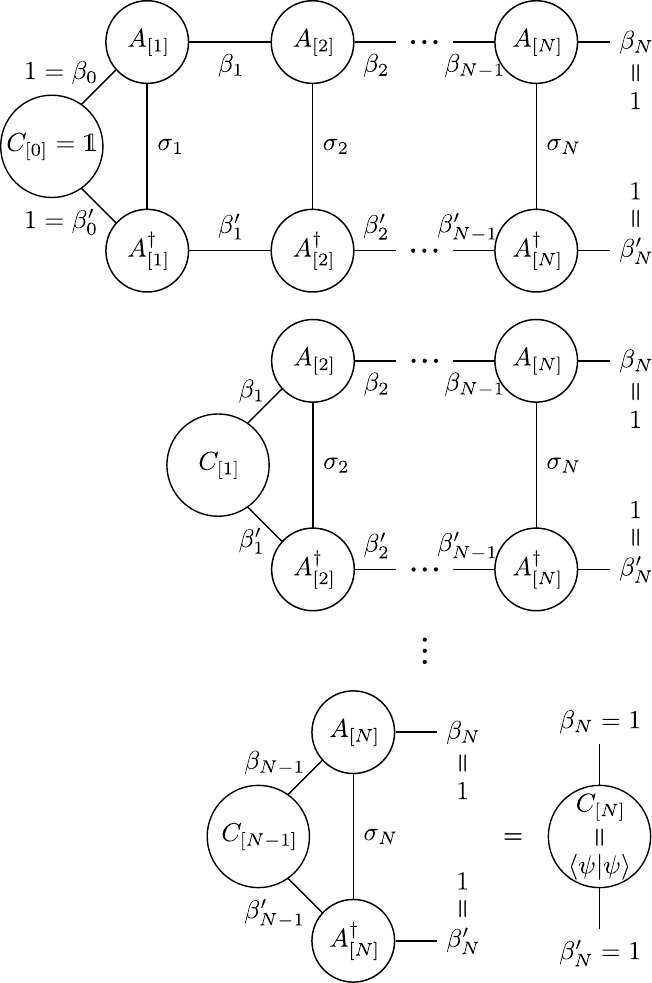}
 \caption{Schematic description of closing-the-zipper strategy to perform contraction of tensor network resulting from the overlap between two MPSs representing the ket $\ket{\psi}$ and the bra $\bra{\psi}$ of a given state to yield $\braket{\psi}$. 
 The steps are ordered from top to bottom. 
 In the first step, $C_{[0]}$ is initialized as the $1 \times 1$ identity matrix and introduced on the left end of the tensor network, being contracted with the leftmost local tensors $A_{[1]}$ and $A^{\dagger}_{[1]}$ through the singleton dummy indices $\beta_0$ and $\beta'_0$. 
 The contraction of the three tensors $C_{[0]}$, $A_{[1]}$ and $A^{\dagger}_{[1]}$---following the strategy described in Fig.~\ref{fig:contractcost}(b)---produces the rank-$2$ tensor $C_{[1]}$. 
 This three-tensor contraction is repeated $N-1$ times until arriving at the final $1 \times 1$ $C_{[N]}$, which is just the desired $\braket{\psi}$. 
 Although this figure considers the computation of the norm of a state $\ket{\psi}$, this scheme can be identically applied to compute the overlap between two distinct MPSs. 
 The closing-the-zipper method can be similarly performed from right to left instead. 
 Assuming the free indices of the MPSs have dimension $d$ and the bond dimension cutoff is $D$, the closing-the-zipper method cost scales as $\mathcal{O}(N D^3 d)$.
 }\label{fig:zippermain}
\end{figure}

In Fig.~\ref{fig:zippermain}, we illustrate this closing-the-zipper contraction scheme of the overlap between two MPSs~\footnote{For simplicity, we consider the computation of the norm, in which case the bra and ket correspond to the same state. 
The generalization to the case of an overlap $\bra{\phi} \ket{\psi}$ between two states $\ket{\phi}$ and $\ket{\psi}$ is straightforward.}. 
The contraction is divided in $N$ steps; at the $n^{\textrm{th}}$ step, the local tensors $A_{[n]}$ and $A^{\dagger}_{[n]}$ are contracted with the tensor $C_{[n-1]}$ that stores the outcome of all contractions from previous steps, yielding the tensor $C_{[n]}$ to be used in the next step:
\vspace{-0.5cm}
\begin{figure}[H]
 \centering
 \includegraphics[width=0.65\linewidth]{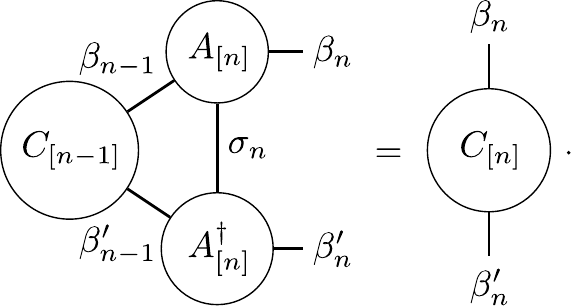}
\end{figure}
\vspace{-0.5cm}
\noindent To make sense of the first and final steps, it is helpful to add singleton dummy indices at each end of the two MPSs, as illustrated in Fig.~\ref{fig:MPSoverlap}(c). 
This allows to apply the first step of the recursive process depicted in Fig.~\ref{fig:zippermain} with $C_{[0]}$ initialized as the $1 \times 1$ identity matrix (i.e., the scalar $1$). 
At the $N^{\textrm{th}}$ and final step, the recursive relation results in the rank-$2$ tensor $C_{[N]}$, with both of its indices $\beta_N$ and $\beta'_N$ having trivial dimension $1$. 
This scalar corresponds precisely to the norm $\bra{\psi} \ket{\psi}$ we were after. 
Of course, we can cover the tensor network from right to left instead, producing exactly the same outcome. 
At each step, we make use of the tensor contraction scheme discussed in Section~\ref{subsec:tensorbasics} (see Fig.~\ref{fig:contractcost}(b)), resulting in a $\mathcal{O}(N D^3 d) \sim \mathcal{O}(N D^3)$ scaling overall. 
Unlike the na\"{i}f approach, the closing-the-zipper strategy allows for a scalable computation of overlaps between MPSs, which is crucial for the practicality of MPS-based DMRG.

\subsubsection{Canonical forms}\label{subsubsec:canonical_forms}

It is possible to cast the MPS in a suitable form that effectively renders most or even all steps of the closing-the-zipper scheme trivial, thus allowing to simplify the tensor network diagrams considerably without requiring any detailed calculations. 
Suppose the MPS is in \textit{left-canonical} form, in which case all local tensors $\{ A_{[i]} \}_{i=1}^{N}$ are left-normalized, i.e.,
\vspace{-0.5cm}
\begin{figure}[H]
 \centering
 \includegraphics[width=0.55\linewidth]{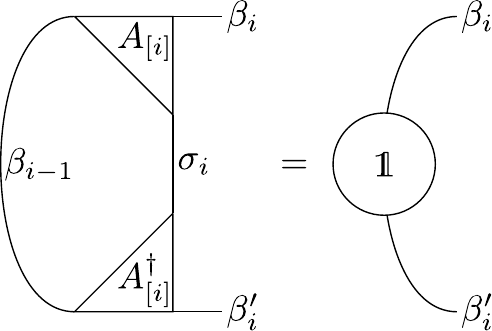}
\end{figure}
\vspace{-0.5cm}
\noindent or $\sum_{\beta_{i-1}, \sigma_i} A^{*}_{\beta'_{i}, \sigma_i, \beta_{i-1}} A_{\beta_{i-1}, \sigma_i, \beta_{i}} = \delta_{\beta'_{i}, \beta_{i}}$ algebraically.
In such case, all $\{ C_{[n]} \}_{n=0}^{N}$ in the closing-the-zipper scheme of Fig.~\ref{fig:zippermain} are just resolutions of the identity, so all steps are trivial and the MPS is normalized, $\braket{\psi} = 1$. 
The same conclusions hold if the closing-the-zipper scheme is performed from right to left and the local tensors are all right-normalized, 
\vspace{-0.5cm}
\begin{figure}[H]
 \centering
 \includegraphics[width=0.65\linewidth]{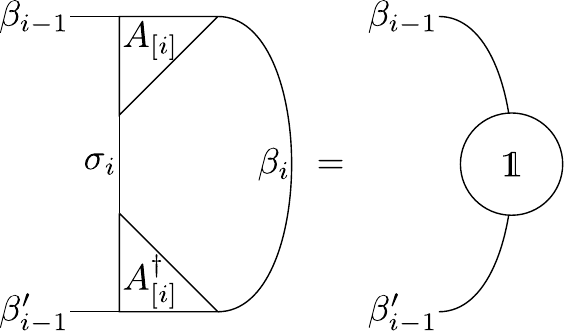}
\end{figure}
\vspace{-0.5cm}
\noindent or $\sum_{\beta_{i}, \sigma_i} A_{\beta_{i-1}, \sigma_i, \beta_{i}} A^{*}_{\beta_{i}, \sigma_i, \beta'_{i-1}} = \delta_{ \beta_{i-1}, \beta'_{i-1}}$ algebraically. 
This is the \textit{right-canonical} form.

For the purposes of computing expectation values of local operators, it is convenient to introduce another canonical form, the so-called \textit{site-canonical} MPS, whereby all local tensors to the left of the site on which the local operator acts nontrivially are left-normalized, and all local tensors to the right are right-normalized. 
To show the usefulness of the site-canonical form, let us consider the expectation value $\bra{\psi} \hat{O}_{[i]} \ket{\psi}$ of a one-site operator (acting on a given site $i$) $\hat{O}_{[i]} = \sum_{\sigma_i, \sigma'_i} O_{\sigma_{i}, \sigma_{i'}} \ket{\sigma_i} \bra{\sigma'_i}$, represented diagrammatically as:
\vspace{-0.5cm}
\begin{figure}[H]
 \centering
 \includegraphics[width=\linewidth]{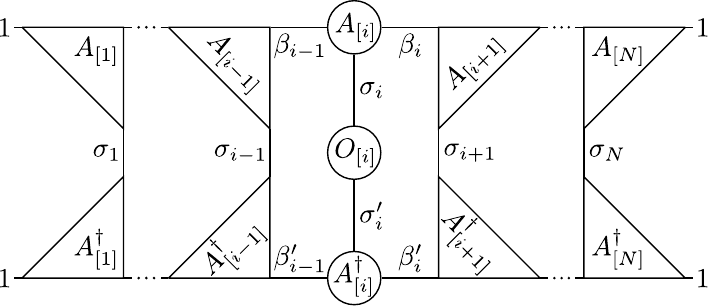}
\end{figure}
\vspace{-0.5cm}
\noindent Upon making use of the left- and right-normalization of the tensors to the left and to the right of site $i$, closing the zipper on either side reduces this one-site-operator expectation value to:
\vspace{-0.5cm}
\begin{figure}[H]
 \centering
 \includegraphics[width=0.4\linewidth]{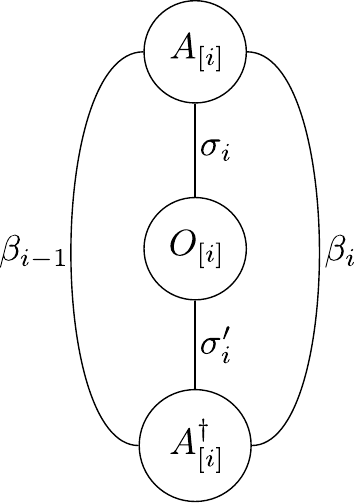}
\end{figure}
\vspace{-0.5cm}

Any MPS can be converted into left-canonical form by performing SVD on one site at a time, covering the full chain from left to right. 
As discussed in the final paragraph of Section~\ref{subsec:svd}, each local rank-$3$ tensor $A_{[i]}$ can be reshaped by fusing the leftmost index with the physical index; the SVD of the resulting matrix yields a unitary matrix $U$, which becomes a left-normalized rank-$3$ tensor upon splitting the two indices that were originally fused. 
Hence, at each site, we replace the original local tensor $A_{[i]}$ with the reshaped $U$, absorbing the remaining $S V^{\dagger}$ in the following local tensor $A_{[i+1]}$. 
At the very last site, because the rightmost index is a singleton dummy index, $V^{\dagger}$ is just a complex number of modulus $1$, so we can neglect it as any wave function is defined up to a global phase factor. 
Moreover, $S$ is a positive real number that corresponds to the norm of the original MPS. 
Typically, $S$ is also discarded, in which case the left-canonical MPS becomes normalized. 
To obtain a right-canonical MPS, one proceeds analogously to the left-canonical case, with the main differences being that the chain is covered from right to left, the local tensor $A_{[i]}$ is replaced by the right-normalized $V^{\dagger}$ resulting from the SVD at that site, and the remaining $U S$ is absorbed by $A_{[i-1]}$. 
For a site-canonical MPS, each of the two processes is carried out on the corresponding side of the selected site.

\subsubsection{General wave function representation}\label{subsubsec:MPS_general-wavefunction}

Being 1D tensor networks, MPSs are most naturally suited for the representation of wave functions of 1D quantum systems. 
However, it should be stressed that any wave function, regardless of its dimensionality or entanglement structure, can be represented as an MPS, though possibly with exceedingly large bond dimensions. 
Suppose we are given the wave function of a quantum system defined on a $N$-site lattice,
\begin{equation}
    \ket{\psi} = \sum_{\sigma_1,\sigma_2,...,\sigma_N} \psi_{\sigma_1,\sigma_2,...,\sigma_N} \ket{\sigma_1} \otimes \ket{\sigma_2} \otimes ... \otimes \ket{\sigma_N},
\end{equation}
where $\ket{\sigma_i}$ denotes the local basis of site $i$.
Assuming the dimension of the local Hilbert space at each site is $d$, the amplitudes of the wave function, $\psi_{\sigma_1,\sigma_2,...,\sigma_N}$, typically cast in the form of a $d^{N}$-dimensional vector, can be reshaped into a rank-$N$ tensor such as the one shown in Fig.~\ref{fig:tensors}(f), with each index having dimension $d$. 
To convert this rank-$N$ tensor into the corresponding $N$-site MPS (Fig.~\ref{fig:tensors}(e)), one can perform SVD at each site at a time following some path that covers every lattice site once~\footnote{In one dimension, the natural choice of path is just to go through the chain from one end to the other. 
In two and higher dimensions, one may consider following a zigzag path that covers one line of each dimension at a time (see Fig.~\ref{fig:2D_1D-long-range} for an example in two dimensions). 
In any case, this conversion of a general rank-$N$ tensor into an $N$-site MPS via a sequence of SVDs works irrespective of the sequence of sites chosen.}. 
At the first site, the original rank-$N$ tensor is reshaped into a $d \times d^{N-1}$ matrix; its SVD produces a unitary $d \times d$ matrix $U$, which is the first local tensor $A_{[1]}$ of the MPS. 
The remainder of the SVD, the $d \times d^{N-1}$ matrix $S V^{\dagger}$, is reshaped into a $d^2 \times d^{N-2}$ matrix, the SVD of which yields a unitary $d^{2} \times d^{2}$ matrix $U$, which is reshaped into a left-normalized rank-$3$ tensor with shape $d \times d \times d^{2}$, corresponding to the second local tensor $A_{[2]}$ of the MPS. 
This sequence of sitewise SVDs is carried out until reaching the last site, where one obtains a rank-$2$ tensor $A_{[N]}$ of dimensions $d \times d$. 
The final outcome is therefore a left-canonical MPS. 
Importantly, because no truncations were performed, until the center of the MPS is reached, the bond dimension keeps on increasing by a factor of $d$ at each site, yielding a maximum bond dimension of $d^{\lfloor N/2 \rfloor}$, which is exponentially large in the system size. 
This is consistent with the fact that no information was lost, so the number of entries of the MPS is $\mathcal{O}(d^N)$, as for the original rank-$N$ tensor. 

The exact conversion of a wave function into an MPS ultimately defeats the purpose of using MPSs (or tensor networks, more generally), which is to provide a more compact representation without compromising the quantitative description of the system under study. 
A more scalable approach would involve truncating the bond dimension of the MPS to a cutoff $D$ set beforehand, although this only produces an approximation of the original state, in general. 
The remarkable success of MPS-based methods in the study of 1D quantum phenomena is rooted upon the favorable scaling of the required bond dimension cutoff $D$ of MPSs with the system size $N$, in accordance with the entanglement area laws discussed in Section~\ref{subsubsec:efficiency}.
The relation between the entanglement entropy of a state in a given bipartition and the corresponding bond dimension $D$ of its MPS representation will be clarified in Section~\ref{sec:5.2}.

\subsection{Matrix product operators}\label{sec:MPO}

\begin{figure}
 \centering
 \includegraphics[width=0.8\linewidth]{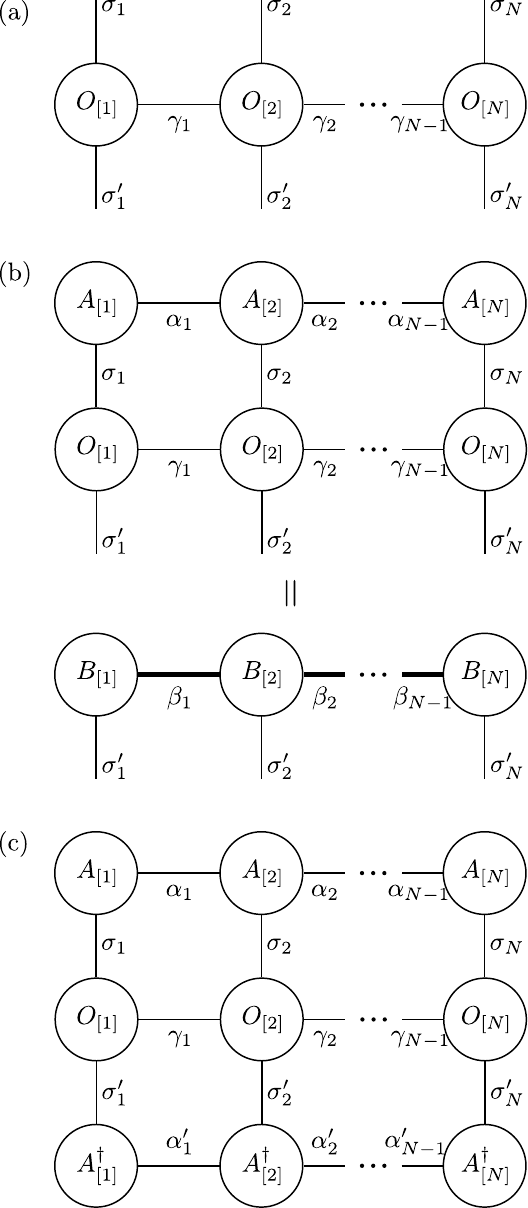}
 \caption{
 (a) Diagram of $N$-site operator $\hat{O}$ as an MPO. 
 (b) Applying MPO of $\hat{O}$ onto MPS of $\ket{\psi}$ yields another MPS. 
 At $m^{\textrm{th}}$ bond, contracted index $\beta_{m}$ of MPS of $\hat{O} \ket{\psi}$ results from fusion of corresponding indices $\gamma_m$ and $\alpha_{m}$ of MPO and original MPS, so the dimension of $\beta_m$ is the product of the dimensions of $\alpha_m$ and $\gamma_m$, hence the bold line representation in the diagram. 
 (c) Diagram of expectation value $\bra{\psi} \hat{O} \ket{\psi}$. 
 The contraction of the tensor network can be done in two ways. 
 Either the MPO is applied to one of the MPSs and then the closing-the-zipper strategy (Fig.~\ref{fig:zippermain}) is adopted to compute the overlap between the two remaining MPSs or the closing-the-zipper method is applied directly to this three-layer tensor network, as described in the text. 
 Assuming the physical indices have dimension $d$ and the bond dimension cutoffs are $D$ for the MPS and $w$ for the MPO, the cost of the two strategies scales as $\mathcal{O}(N D^3 w^2 d)$ and $\mathcal{O}(N D^3 w d)$, respectively.
 }
 \label{fig:MPO_MPS_contractions}
\end{figure}

A \textit{matrix product operator} (MPO) is a 1D tensor network of the form shown diagrammatically in Fig.~\ref{fig:MPO_MPS_contractions}(a). 
The structure of an MPO is similar to that of an MPS, except for the number of physical indices. 
While an MPS has a single physical index per site, an MPO has two, the top one to act on kets and the bottom one to act on bras, following the convention adopted in Fig.~\ref{fig:MPSoverlap}. 
MPOs constitute the most convenient representation of operators for MPS-based methods, as they allow for a sitewise update of the MPS ansatz. 
In particular, the MPS-based formulation of DMRG discussed in Section~\ref{sec:finite-DMRG_TN} involves expressing the Hamiltonian under study as an MPO.

Applying an MPO onto an MPS yields another MPS of greater bond dimensions (Fig.~\ref{fig:MPO_MPS_contractions}(b)). 
To obtain this MPS, at every site $i = 1, 2, ..., N$ one contracts the local tensor $A_{[i]}$ from the original MPS with the corresponding local tensor $O_{[i]}$ from the MPO, fusing the pairs of bonds on either side to retrieve a rank-$3$ tensor $B_{[i]}$,
\vspace{-0.5cm}
\begin{figure}[H]
 \centering
 \includegraphics[width=0.9\linewidth]{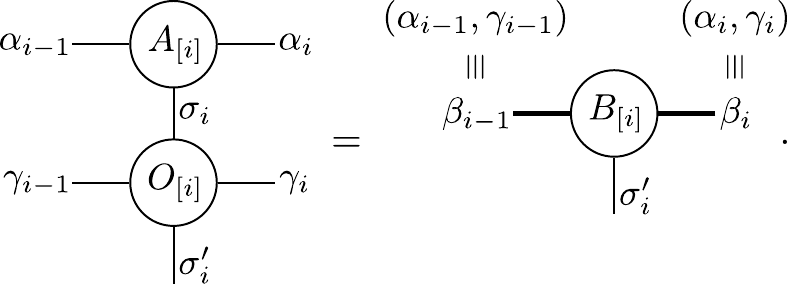}
\end{figure}
\vspace{-0.5cm}
\noindent Due to this fusion of indices, the bond dimensions of the final MPS are the product of the bond dimensions of the original MPS and the MPO. 
The cost of contracting an MPO of bond dimension $w$ (which is typically a small constant, as we shall see below for the case of a short-ranged Hamiltonian) with an MPS of bond dimension $D$, both with $N$ sites and physical index dimension $d$, is $\mathcal{O}(N D^2 w^2 d^2) \sim \mathcal{O}(N D^2)$.

The expectation value of an operator $\hat{O}$ cast in the form of an $N$-site MPO with respect to a state $\ket{\psi}$ expressed as a $N$-site MPS is represented in Fig.~\ref{fig:MPO_MPS_contractions}(c). 
One way to obtain $\bra{\psi} \hat{O} \ket{\psi}$ is to calculate the MPS corresponding to $\hat{O} \ket{\psi}$---following the prescription provided in the previous paragraph---and then compute the overlap between the two resulting MPSs, one for $\hat{O} \ket{\psi}$ and another for $\bra{\psi}$, using the closing-the-zipper method introduced in Section~\ref{subsec:MPSoverlaps}. 
The cost of this approach scales as $\mathcal{O}(N D^3 w^2 d) \sim \mathcal{O}(N D^3)$. 
Alternatively, the closing-the-zipper strategy can be adapted to contract this three-layer tensor network. 
Specifically, at the $n^{\textrm{th}}$ iteration we have
\vspace{-0.5cm}
\begin{figure}[H]
 \centering
 \includegraphics[width=0.8\linewidth]{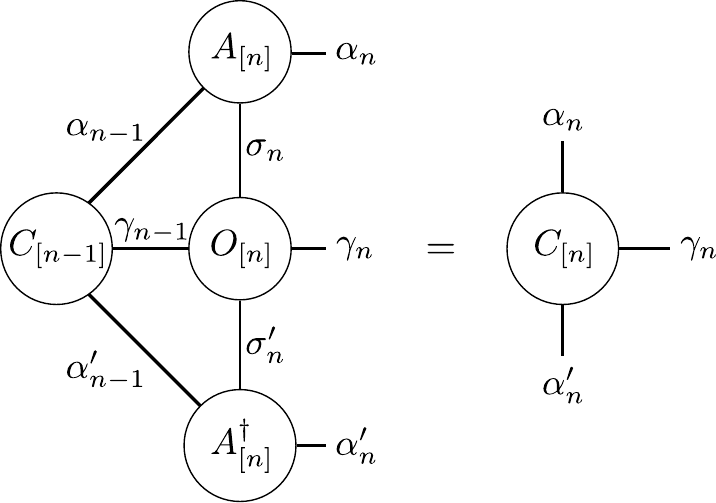}
\end{figure}
\vspace{-0.5cm}
\noindent and the indices are contracted as follows:
\begin{enumerate}
    \item{Sum over $\alpha_{n-1}$ with fixed $\sigma_n$, $\alpha_n$, $\gamma_{n-1}$ and $\alpha'_{n-1}$ at cost $\mathcal{O}(D^3 w d)$.}
    
    \item{Sum over $\gamma_{n-1}$ and $\sigma_n$ with fixed $\alpha_n$, $\gamma_n$, $\sigma'_{n}$ and $\alpha'_{n-1}$ at cost $\mathcal{O}(D^2 w^2 d^2)$.}
    
    \item{Sum over $\alpha'_{n-1}$ and $\sigma'_n$ with fixed $\alpha_n$, $\gamma_n$ and $\alpha'_{n}$ at cost $\mathcal{O}(D^3 w d)$.}
\end{enumerate}
Upon completing the $N$ iterations to go through all sites, the overall scaling is $\mathcal{O}(N D^3 w d) \sim \mathcal{O}(N D^3)$.
For technical reasons that will be apparent in Section~\ref{sec:5.1}, this contraction scheme is preferred in the implementation of the finite-system DMRG algorithm.

Any $N$-site operator can be expressed as an MPO by performing SVD at each site at a time, in a similar spirit to the representation of an arbitrary wave function in terms of an MPS, discussed in Section~\ref{subsubsec:MPS_general-wavefunction}.
The problem with this approach is that the bond dimension of the resulting MPO grows by $d^2$ at every iteration until reaching the middle of the MPO, thus leading to $\mathcal{O}(d^N)$ bond dimensions. 
The MPO representation of an arbitrary tensor product of single-site operators is straightforward: each local operator is reshaped into a rank-$4$ tensor with two singleton dummy indices (corresponding to the trivial bonds with dimension $w = 1$), which are contracted with those from the adjacent sites to form the MPO.
MPOs like those described above can also be summed~\footnote{See, e.g., Ref.~\cite{Hubig2017} for a general prescription, which amounts to writing each rank-$4$ local tensor of the MPOs that we want to sum as a matrix of the physical operators, and then perform direct sums of these matrices at every site, except for the leftmost/rightmost site where the physical operators are organized in a line/column vector.} in order to obtain the MPO representation of more generic operators.
It must be noted, however, that the previous strategy, although versatile, does not always lead to the lowest possible bond dimensions of the final MPO.
In particular, it is possible to represent local Hamiltonians in terms of MPOs with $\mathcal{O}(1)$ bond dimension---i.e., constant with respect to the system size $N$---, as explained below.

The exact MPO of a local Hamiltonian can be obtained through an analytical method originally proposed by McCulloch~\cite{McCulloch2007}. 
For concreteness, let us consider the Heisenberg model for an open-ended spin-$s$ chain with a Zeeman term,
\begin{align}
\hat{\mathcal{H}} &= J \sum_{i = 1}^{N-1} \hat{\vec{S}}_{i} \cdot \hat{\vec{S}}_{i+1}  - h \sum_{i = 1}^{N} \hat{S}^{z}_{i} \\
&= J \sum_{i = 1}^{N-1} \left( \hat{S}_{i}^{z} \hat{S}_{i+1}^{z} + \frac{\hat{S}_{i}^{+} \hat{S}_{i+1}^{-} + \hat{S}_{i}^{-} \hat{S}_{i+1}^{+}}{2} \right) \nonumber \\
& \quad - h \sum_{i = 1}^{N} \hat{S}^{z}_{i},
\label{eq:Hamiltonian_MPO}
\end{align}
where $J$ and $h$ are model parameters, $\hat{\vec{S}}_{i} = \left(\hat{S}^{x}_{i},\hat{S}^{y}_{i},\hat{S}^{z}_{i}\right)$ is the vector of spin-$s$ operators at site $i \in \{1,2,...,N\}$, and $\hat{S}_{i}^{\pm} = \hat{S}_{i}^{x} \pm \mathrm{i} \hat{S}_{i}^{y}$ are the corresponding spin ladder operators.
Our goal is to obtain the local tensors $\{H_{[i]}\}_{i=1}^{N}$ of the MPO that encodes this Hamiltonian. 
Four different types of terms arise in Eq.~\eqref{eq:Hamiltonian_MPO}:
\begin{equation*}
    \begin{aligned}
    & ... \stackrel{5}{\otimes} \hat{\mathds{1}} \stackrel{5}{\otimes} J \hat{S}^{z} \; \; \, \stackrel{2}{\otimes} \hat{S}^{z} \, \stackrel{1}{\otimes} \hat{\mathds{1}} \stackrel{1}{\otimes} ... \\
    & ... \stackrel{5}{\otimes} \hat{\mathds{1}} \stackrel{5}{\otimes} \frac{J}{2} \hat{S}^{+} \; \stackrel{3}{\otimes} \hat{S}^{-} \stackrel{1}{\otimes} \hat{\mathds{1}} \stackrel{1}{\otimes} ... \\
    & ... \stackrel{5}{\otimes} \hat{\mathds{1}} \stackrel{5}{\otimes} \frac{J}{2} \hat{S}^{-} \; \stackrel{4}{\otimes} \hat{S}^{+} \stackrel{1}{\otimes} \hat{\mathds{1}} \stackrel{1}{\otimes} ... \\
    & ... \stackrel{5}{\otimes} \hat{\mathds{1}} \stackrel{5}{\otimes} -h \hat{S}^{z} \stackrel{1}{\otimes} \hat{\mathds{1}} \; \; \, \stackrel{1}{\otimes} \hat{\mathds{1}} \stackrel{1}{\otimes} ...
    \end{aligned}
\end{equation*}
\noindent The numbers above the tensor product signs identify one of the following five `states':
\begin{itemize}
    \item{`State' 1: Only identity operators $\hat{\mathds{1}}$ to the right.}
    \item{`State' 2: One $\hat{S}^{z}$ operator just to the right, followed by $\hat{\mathds{1}}$ operators.}
    \item{`State' 3: One $\hat{S}^{-}$ operator just to the right, followed by $\hat{\mathds{1}}$ operators.}
    \item{`State' 4: One $\hat{S}^{+}$ operator just to the right, followed by $\hat{\mathds{1}}$ operators.}
    \item{`State' 5: One complete term somewhere to the right.}
\end{itemize}
For a given bulk site $i$, the local rank-$4$ tensor $H_{[i]}$, cast in the form of a $w \times w$ matrix where each entry is itself a $d \times d$ matrix---with $w=5$ the bond dimension of the MPO (determined by the number of `states') and $d = 2s + 1$ the physical index dimension---, is constructed in such a way that its $(k,l)$ entry corresponds to the operator that makes the transition from `state' $l$ to `state' $k$ towards the left:
\begin{equation}
    H_{[i]} = 
    \begin{pmatrix}
    \hat{\mathds{1}}_i         & 0 & 0 & 0 & 0 \\
    \hat{S}_{i}^{z}    & 0 & 0 & 0 & 0 \\
    \hat{S}_{i}^{-}    & 0 & 0 & 0 & 0 \\
    \hat{S}_{i}^{+}    & 0 & 0 & 0 & 0 \\
    -h \hat{S}_{i}^{z} & J \hat{S}_{i}^{z} & \frac{J}{2} \hat{S}_{i}^{+} & \frac{J}{2} \hat{S}_{i}^{-} & \hat{\mathds{1}}_i
    \end{pmatrix}.
    \label{eq:MPO_Heisenberg}
\end{equation}
For the terminal sites, due to the open boundary conditions, we have two rank-$3$ tensors, one corresponding to the last row of Eq.~\eqref{eq:MPO_Heisenberg} for the leftmost site $i = 1$ and another corresponding to the first column of Eq.~\eqref{eq:MPO_Heisenberg} for the rightmost site $i = N$.

In order to confirm that the constructed MPO does indeed give rise to the Hamiltonian stated in Eq.~\eqref{eq:Hamiltonian_MPO}, one can perform by hand the matrix multiplication of the local tensors in the form shown in Eq.~\eqref{eq:MPO_Heisenberg}, but with the usual scalar multiplications being replaced by tensor products as each entry is itself a rank-$2$ tensor~\cite{CrosswhiteBacon2008}. 
Alternatively, the MPO can be contracted and compared directly to the full $d^{N} \times d^{N}$ matrix representation of the model Hamiltonian. 
This sanity check is performed for small system sizes $N$ in the code that complements this manuscript (see Supplementary Information). 
In this code, we also construct the MPO Hamiltonian for two other quantum spin models, the Majumdar--Ghosh~\cite{Majumdar1969,Majumdar1969a} and the Affleck--Kennedy--Lieb--Tasaki~\cite{Affleck1987} models. 
These two additional examples suffice to demonstrate how to apply McCulloch's method in general, namely by adding next-nearest-neighbor interactions and further nearest-neighbor interactions, respectively. 
Assuming the most conventional case of model Hamiltonians with terms acting nontrivially on one or two sites only, the bond dimension of the MPO obtained with this method starts at two and increases by one for every new type of two-site term and/or unit of interaction range~\cite{Schollwock2011}.
There are, however, notable exceptions to this rule, such as long-range Hamiltonians that allow for a more compact but still exact MPO representation~\cite{Crosswhite2008,Froewis2010}. 
More complex Hamiltonians such as those arising in quantum chemistry~\cite{Chan2011} or in two-dimensional lattice models on a cylinder in hybrid real and momentum space~\cite{Motruk2016} may require more sophisticated numerical approaches to reduce the bond dimension of the corresponding MPO~\cite{Hubig2017}. 

\section{Finite-system DMRG in the language of tensor networks}\label{sec:finite-DMRG_TN}

\subsection{Derivation: one-site update} \label{sec:5.1}

\begin{figure*}
 \centering
 \includegraphics[width=0.85\linewidth]{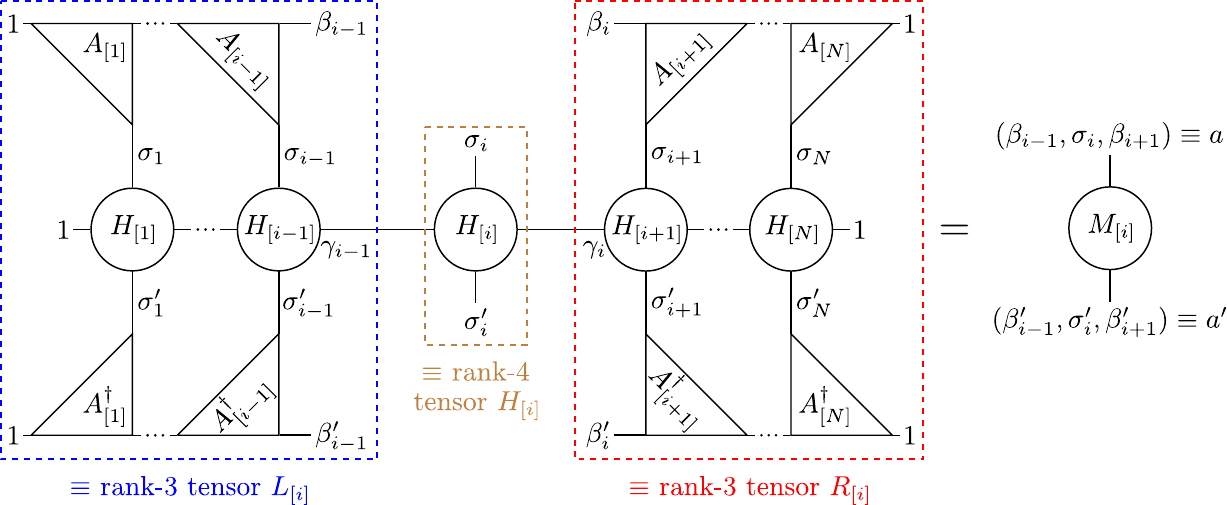}
 \caption{
 Tensor network diagram of the ``effective'' matrix $M_{[i]}$ of the eigenvalue problem (Eq.~\eqref{eq:eigenvalue_eq}) associated with one iteration---corresponding to the local optimization of the MPS at site $i$---of the one-site-update finite-system DMRG algorithm.
 To optimize the computational performance of the DMRG algorithm, $M_{[i]}$ is stored in terms of three tensors, $L_{[i]}$, $H_{[i]}$ and $R_{[i]}$ (see Appendix~\ref{app:L_R_update} for details).
 }
 \label{fig:Fig_15}
\end{figure*}

The starting point for the derivation of the MPS-based finite-system DMRG algorithm is to consider the set of all $N$-site MPS representations of a ket $\ket{\psi}$ with (maximum) bond dimension $D$ as a variational space.
The local tensor of the MPS at site $i$ is denoted by $A_{[i]}$; for the sake of simplicity, we consider that the physical index dimension is $d$ at all sites.
We assume we are given the $N$-site MPO representation of the Hamiltonian $\hat{\mathcal{H}}$, with bond dimension $w \sim \mathcal{O}(1)$ and physical index dimension $d$; its local rank-$4$ tensor at site $i$ is denoted by $H_{[i]}$. 
The goal is to minimize the energy $\bra{\psi} \hat{\mathcal{H}} \ket{\psi}$, subject to the normalization constraint $\braket{\psi} = 1$.
This can be achieved by minimizing the cost function $\bra{\psi} \hat{\mathcal{H}} \ket{\psi} - \lambda \braket{\psi}$, where $\lambda$ denotes the Lagrange multiplier.
The one-site-update version of the algorithm consists of finding the stationary points of the cost function with respect to each local tensor $A^\dagger_{[i]}$ at a time,
i.e.,
\begin{equation}
    \frac{\partial}{\partial A^\dagger_{[i]}} \left( \bra{\psi} \hat{\mathcal{H}} \ket{\psi} - \lambda \braket{\psi} \right) = 0.
\label{eq:optimization_problem}
\end{equation}
Making use of the diagrammatic representation, and taking into account that all contractions on a tensor network are linear operations, the derivative with respect to $A^\dagger_{[i]}$ amounts to \textit{punching a hole}~\cite{vonDelft2020} at the position of the tensor $A^\dagger_{[i]}$, leading to
\vspace{-0.5cm}
\begin{figure}[H]
 \centering
 \includegraphics[width=\linewidth]{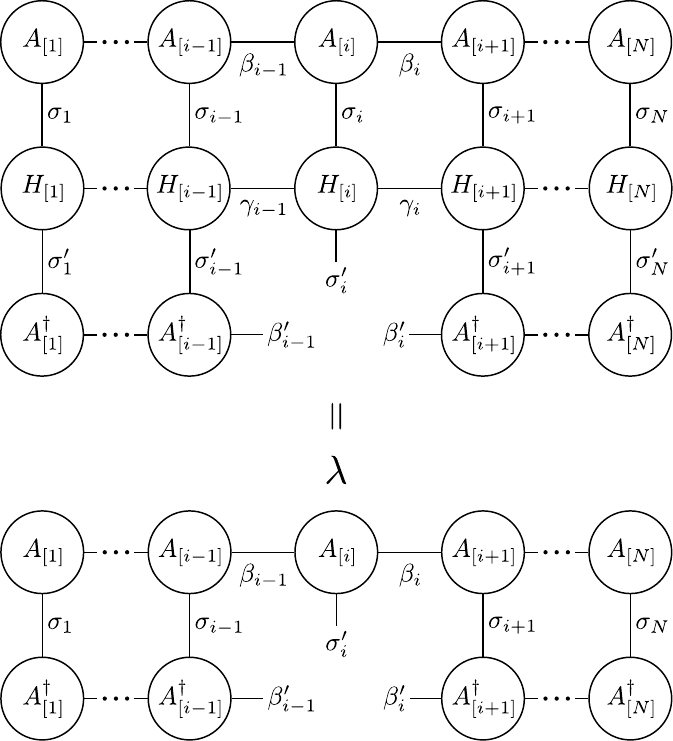}
\end{figure}
\vspace{-0.5cm}
\noindent which can be understood as a generalized eigenvalue problem for $A_{[i]}$.
By casting the MPS in site-canonical form with respect to site $i$, the bottom part of the previous equation simplifies trivially, yielding an eigenvalue problem for $A_{[i]}$ that we write as
\begin{equation}
    \sum_{a} M^{a',a}_{[i]} A^a_{[i]} = \lambda A^{a'}_{[i]}, 
    \label{eq:eigenvalue_eq}
\end{equation}
with $a \equiv (\beta_{i-1},\sigma_i,\beta_i)$ and $M^{a',a}_{[i]}$ defined by the diagram shown in Fig.~\ref{fig:Fig_15}.

Having derived an eigenvalue problem (Eq.~\eqref{eq:eigenvalue_eq}) from the local optimization of the MPS at site $i$ (Eq.~\eqref{eq:optimization_problem}), the optimal update of the corresponding local tensor $A_{[i]}$ is simply the eigenstate with lowest eigenvalue, both of which can be obtained through the Lanczos algorithm~\cite{Lanczos1950}.
In addition to the obtained eigenstate being the variationally optimized $A_{[i]}$, the corresponding eigenvalue is also the current estimate of the ground state energy of the full system.
This step of the DMRG algorithm is repeated, sweeping $i$ back and forth between $1$ and $N$.
As for the initialization, the typical approach is to start with a random MPS.

Two additional technical remarks regarding the implementation of the DMRG algorithm derived above are in order. 
First, at every step of the algorithm, after having obtained the updated local tensor $A_{[i]}$ as the ground state of the eigenvalue problem, its SVD is performed to ensure that the MPS is in the appropriate site-canonical form in the next step of the sweep, thus avoiding a generalized eigenvalue equation.
Second, the ``effective'' matrix of the eigenvalue problem, $M_{[i]}$, is stored in terms of three separate tensors, $L_{[i]}$, $H_{[i]}$ and $R_{[i]}$ (Fig.~\ref{fig:Fig_15}). 
As the notation suggests, the rank-$4$ tensor $H_{[i]}$ is just the local tensor at site $i$ of the MPO that encodes the Hamiltonian $\hat{\mathcal{H}}$. 
As for the rank-$3$ tensors $L_{[i]}$ and $R_{[i]}$, they result from the contraction of all tensors to the left and to the right of site $i$, respectively. 
The efficient computation of $L_{[i]}$ and $R_{[i]}$ over the multiple sweeps of the DMRG algorithm is detailed in Appendix~\ref{app:L_R_update}.

Making use of the internal structure of the matrix $M_{[i]}$, the time complexity of solving the eigenvalue problem stated in Eq.~\eqref{eq:eigenvalue_eq}---required to update one local tensor of the MPS---is $\mathcal{O}(D^3)$. 
This scaling results largely from the the matrix-vector multiplications involved in the construction of the Krylov space within the Lanczos algorithm~\cite{Koch2011}. 
Note that the na\"{i}f explicit contraction of $M_{[i]}$ into a $(D^2 d) \times (D^2 d)$ matrix would have resulted in a $\mathcal{O}(D^4)$ scaling of the matrix-vector multiplications, as opposed to the $\mathcal{O}(D^3)$ obtained using the $L_{[i]}$, $H_{[i]}$ and $R_{[i]}$ tensors.
In the end, all key steps of one iteration of the one-site-update finite-system DMRG algorithm---closing-the-zipper contraction (as described in Appendix~\ref{app:L_R_update}), eigenvalue problem, and SVD---have the same $\mathcal{O}(D^3)$ computational cost, so the overall cost of a full sweep scales as $\mathcal{O}(N D^3)$.
It must be noted that, since the standard Python functions to implement the Lanczos algorithm (e.g., \texttt{scipy.sparse.linalg.eigsh}) require a matrix as input, the na\"{i}f explicit contraction of $M_{[i]}$ was adopted in the code that complements this manuscript, trading efficiency for simplicity.

Finally, although this discussion has been restricted to the computation of the ground state, it is straightforward to extend it to the calculation of low-lying excited states. 
For concreteness, let us suppose we have already determined the ground state $\ket{\textrm{GS}}$ in a previous run of the DMRG algorithm and wish to obtain the first excited state.
Exploiting the orthogonality of the eigenbasis of the Hamiltonian, we merely have to impose the additional constraint $\bra{\psi} \ket{\textrm{GS}} = 0$ through another Lagrange multiplier in the cost function.
This additional term effectively imposes an energy penalty on the variational states $\ket{\psi}$ that have nonzero overlap with $\ket{\textrm{GS}}$. 
In other words, the eigenvalue problem is restricted to a subspace orthogonal to $\ket{\textrm{GS}}$. 
In practice, this condition can be imposed by setting $\ket{\textrm{GS}}$ as the first Krylov state in the Lanczos algorithm but performing the diagonalization of the tridiagonal matrix defined in the Krylov subspace spanned by all but the first Krylov state~\cite{Koch2011}.

\subsection{Connection to original formalism} \label{sec:5.2}

The one-site-update MPS-based DMRG algorithm derived in the previous section is entirely analogous to the original formulation of the finite-system DMRG scheme (recall Section~\ref{sec:finite-DMRG}) provided that there is only one site---denoted by $\circ$---, instead of two, between the blocks S and E (adopting the notation employed in Fig.~\ref{fig:finite-DMRG}).
Considering a left-to-right sweep~\footnote{An analogous reasoning is straightforward for the case of a right-to-left sweep.} of the MPS-based formulation, the SVD of the optimized local tensor $A_{[i]}$ at the site $i$ between S and E---which leaves a left-normalized tensor at site $i$ in the MPS representation of the target eigenstate $\ket{\psi}$---and the subsequent contractions to update the $L_{[i+1]}$ tensor (as defined in Fig.~\ref{fig:Fig_15}) correspond to the projection of the Hilbert space of the growing block $\text{S} \circ$ onto the subspace spanned by the highest-weight eigenstates of the reduced density matrix $\sigma_{\text{S} \circ} = \textrm{Tr}_{\text{E}}(\ket{\psi} \bra{\psi})$ considered in the original formulation.
The number of kept eigenstates of the reduced density matrix $\sigma_{\text{S} \circ}$ in the original formulation is precisely the number of kept singular values in the SVD of the optimized local tensor in the MPS-based version, which translates into the bond dimension $D$ of the MPS ansatz. 
To support the previous claims, we note that the eigenvalues of $\sigma_{\text{S} \circ}$ in the original formulation are the square of the singular values $\{ s_{n} \}_{n=1}^{D}$ of the SVD of the updated $A_{[i]}$ in the MPS-based version (see Appendix~\ref{app:no_trunc_original_finite_DMRG} for the derivation),
\begin{align}
    \sigma_{\text{S} \circ} = \textrm{Tr}_{\text{E}}(\ket{\psi} \bra{\psi}) = \sum_{n = 1}^{D} s_{n}^{2} \ket{u_{n}}_{\text{S} \circ} \prescript{}{\text{S} \circ}{\bra{u_{n}}},
\end{align}
where $ \{ \ket{u_{n}}_{\text{S} \circ} \}_{n=1}^{D}$ are the $D$---out of the total $Dd$---eigenstates of $\sigma_{\text{S} \circ}$ with (possibly) nonzero eigenvalues.
Therefore, we see that in both formulations of DMRG, $D$ quantifies the degree of entanglement that can be captured across the bipartition between $\text{S} \circ$ and E.
Moreover, it becomes apparent that the truncation prescribed in the original formulation of the one-site-update finite-system DMRG scheme is actually trivial (see Appendix~\ref{app:no_trunc_original_finite_DMRG} for a more detailed explanation), thus sorting out the apparent contradiction related with the fact that no truncation is prescribed in the MPS-based version of the one-site-update DMRG algorithm.

Although the original and the MPS-based formulations of DMRG are equivalent, there is one key difference between them regarding the encoding of the Hamiltonian. 
While in the original method the Hamiltonian obtained from the prior implementation of infinite-system DMRG is inherently approximate as its matrix representation results from an explicit truncation of the Hilbert space through a projection onto a smaller subspace defined by the highest-weight eigenstates of the reduced density matrices on either side of the bipartition considered, in the MPS-based version the MPO representation of the Hamiltonian with which one begins to perform the sweeps is exact and the approximate description of the system is entirely restricted to the ansatz of the variational problem, an MPS with given bond dimension $D$. 
This difference renders the MPS-based formulation of DMRG more effective at calculating physical quantities related to powers of the Hamiltonian, such as the energy variance or, more generally, cumulant expansions~\cite{Hubig2017}.

Similarly to the original formulation of the finite-system DMRG scheme described in Section~\ref{sec:finite-DMRG}, there is a two-site-update version of MPS-based DMRG that results from simultaneously minimizing the cost function $\bra{\psi} \hat{\mathcal{H}} \ket{\psi} - \lambda \braket{\psi}$ (recall Eq.~\eqref{eq:optimization_problem} and the corresponding derivation) with respect to two adjacent local tensors $A^\dagger_{[i]}$ and $A^\dagger_{[i+1]}$, giving rise to an eigenvalue problem for a two-site tensor of the form
\vspace{-0.5cm}
\begin{figure}[H]
 \centering
 \includegraphics[width=0.95\linewidth]{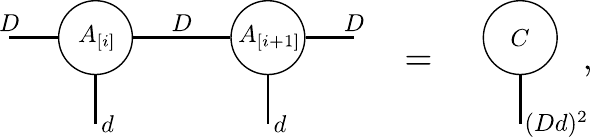}
\end{figure}
\vspace{-0.5cm}
\noindent where the explicit contraction (and index fusion) that casts the two-site tensor in the vectorial form $C$ is not carried out in practice (as discussed in Section~\ref{sec:5.1}) but is nonetheless a useful picture to have in mind. 
For the sake of clarity, the labels associated with the legs represent the dimensions of the corresponding indices. 
Once the eigenvalue problem is solved, the updated tensor $C$ is reshaped into a $(Dd) \times (Dd)$ matrix so that its SVD can be performed to obtain the optimized local tensors at sites $i$ and $i+1$.
Crucially, the MPS bond dimension between the local tensors at sites $i$ and $i+1$ increases from $D$ to $D d$ after this optimization process, so an explicit truncation that keeps only the the $D$ highest singular values is required.
Hence, the two-site-update algorithm effectively surveys a larger search space than the one-site-update scheme. 
In particular, this allows to escape local minima in the optimization landscape, namely by having the possibility to explore different symmetry sectors. 
This is the main reason why the two-site-update DMRG scheme, both in its original and MPS-based formulations, is the standard option in the literature. 
It should be stressed, however, that the one-site-update DMRG algorithm can be just as reliable as the two-site-update scheme at a lower computational cost provided that one adopts a correction to the one-site-update proposed by White in 2005~\cite{White2005}, which introduces quantum fluctuations that effectively avoid remaining stuck in metastable configurations.

In the original formulation of DMRG, the outcome of the infinite-system version is the natural starting point for finite-system DMRG. 
In the MPS-based version, however, it is common practice to start from a random MPS of given bond dimension $D$, although this is not usually as good an educated guess as the outcome of the infinite-system DMRG~\cite{Schollwock2011}. 
Alternatively, one may perform finite-system DMRG simulations with MPSs of progressively larger bond dimension $D$, using the outcome of the previous simulation as the initial state for the current one, padding the remainder of the local tensors with zeros due to the larger bond dimension. 
A particularly elegant aspect of MPS-based DMRG, notably in its one-site-update scheme, is that the manifold of states explored in the variational problem corresponds to all MPSs of fixed bond dimension $D$, with no truncations being performed throughout the computation. 
The tangent-space methods developed in recent years~\cite{Haegeman2013,Haegeman2016} explore this feature in more sophisticated ways.

\subsection{Code implementation}

In the same spirit of Section~\ref{subsubsec:original-DMRG_code}, we provide a practical implementation of an MPS-based DMRG algorithm.
Our code, available both in Supplementary Information and at \url{https://github.com/GCatarina/DMRG_MPS_didactic}, consists of a documented Jupyter notebook, written in Python, that goes through all the key steps required to implement the finite-system DMRG method.
For simplicity, we consider the one-site update version (see Section~\ref{sec:5.1}), targeting only the ground state properties.
It must also be noted that the coded DMRG routine is model-agnostic, requiring only as input an Hamiltonian MPO with trivial leftmost and rightmost legs.
In general, the previous requirement is naturally fulfilled for systems with open boundary conditions in at least one of their physical dimensions.

\begin{figure}
 \centering
 \includegraphics[width=\columnwidth]{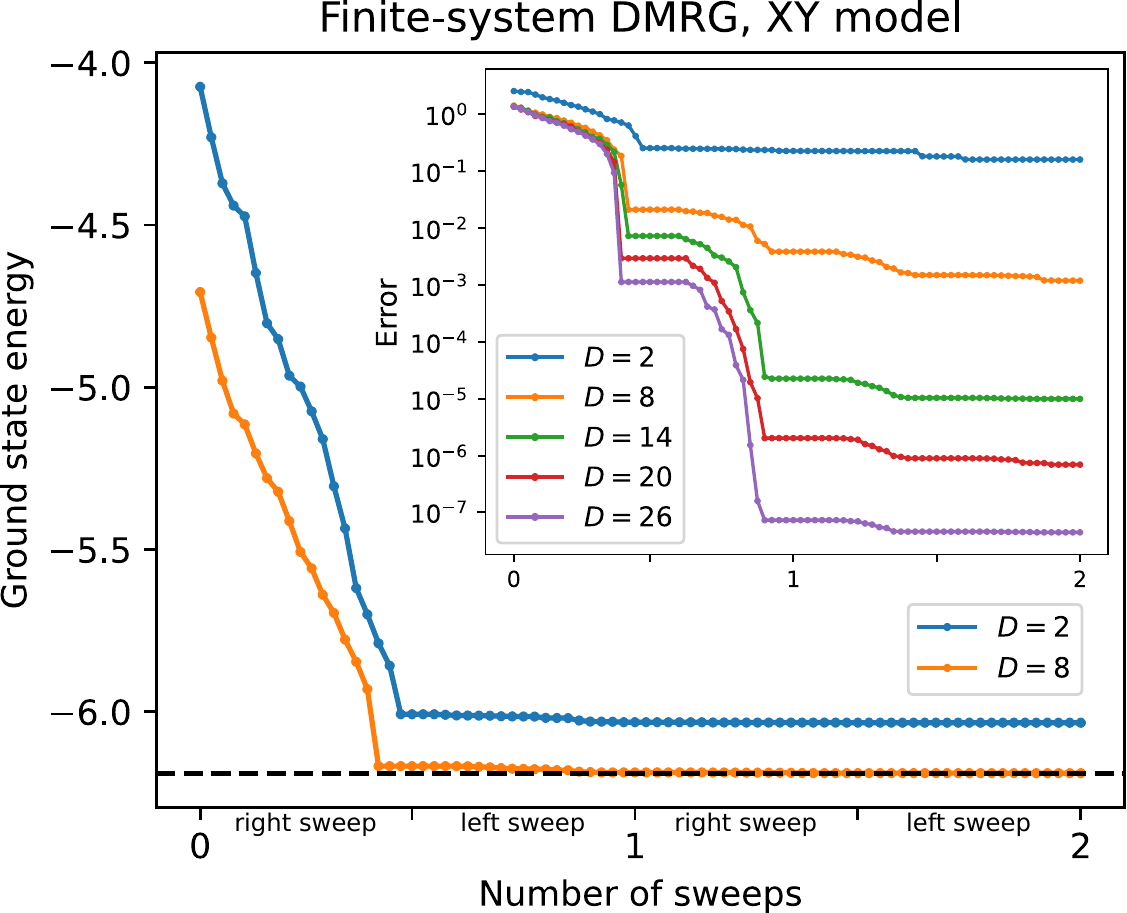}
 \caption{
Finite-system DMRG applied to the (isotropic) XY 1D quantum model.
Ground state energy of an open-ended chain composed of $20$ $s=1/2$ spins, as a function of the number of sweeps of the one-site-update DMRG routine, for different values of the bond dimension cutoff $D$.
The dashed black line marks the analytical result~\cite{Fagotti2012}.
The deviation from the exact solution is also shown in the inset.
 }
 \label{fig:fDMRG_XY_Egs}
\end{figure}

\begin{figure}
 \centering
 \includegraphics[width=\columnwidth]{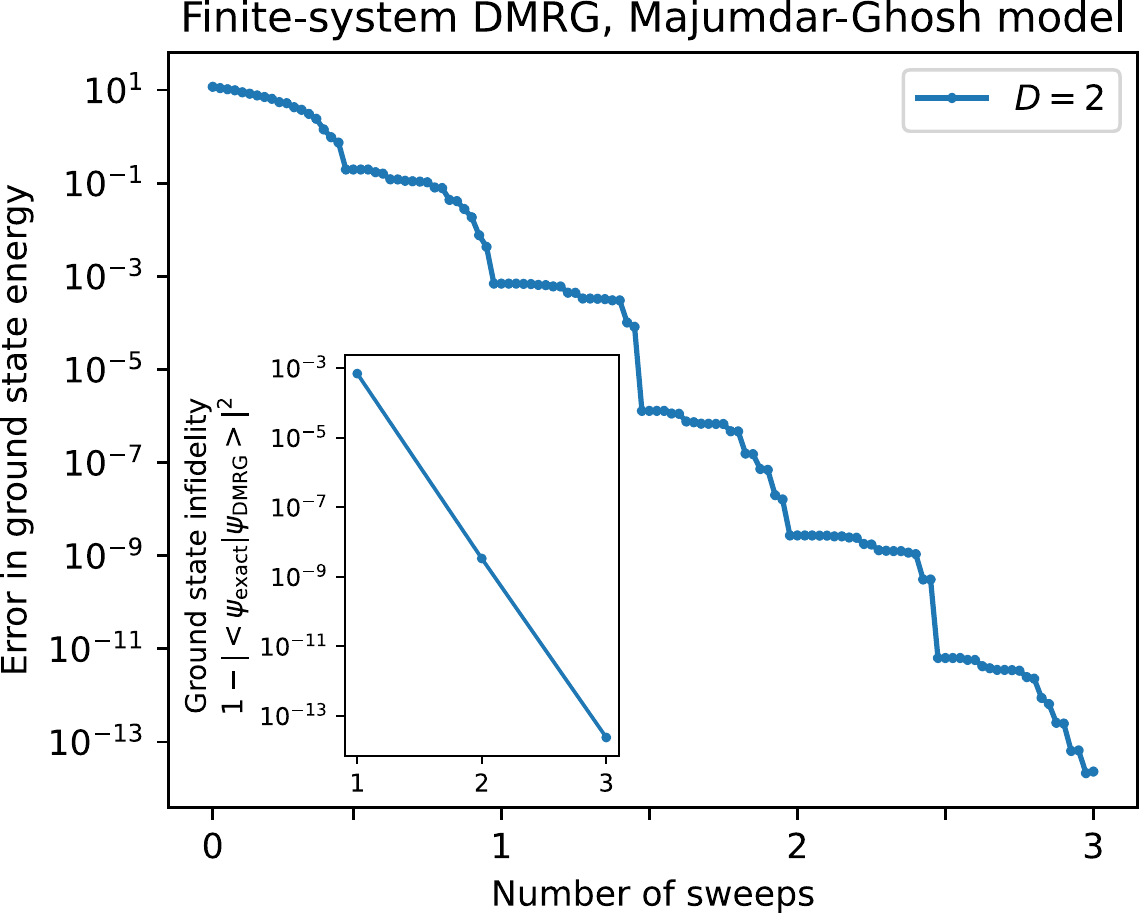}
 \caption{
Finite-system DMRG calculations for the ground state of an open-ended Majumdar--Ghosh chain, whose analytical solution~\cite{Majumdar1969,Majumdar1969a} can be written as an MPS with bond dimension $2$.
Numerical results, obtained with bond dimension cutoff $D=2$, for a chain composed of $20$ $s=1/2$ spins, show the convergence to the exact ground state as a function of the number of sweeps of the one-site-update DMRG algorithm.
 }
 \label{fig:fDMRG_MG_GS}
\end{figure}

In order to benchmark our DMRG code, we apply it to 1D systems, with open boundary conditions, for which exact ground state solutions are known.
Specifically, we consider the (isotropic) XY~\cite{Lieb1961} and the Majumdar--Ghosh~\cite{Majumdar1969,Majumdar1969a} models.
In Fig.~\ref{fig:fDMRG_XY_Egs}, we show how, for an XY chain, the ground state energy computed by DMRG compares with the analytical result~\cite{Fagotti2012}.
It is apparent that DMRG converges rapidly with the number of sweeps performed.
It is also observed that the accuracy of the numerical calculation is determined by the bond dimension cutoff $D$.
In Fig.~\ref{fig:fDMRG_MG_GS}, a complementary example is shown, where DMRG is used to compute both the error in the energy estimation and the infidelity associated with the ground state of a Majumdar--Ghosh chain.
Since, for open-ended Majumdar--Ghosh chains, the exact ground state wave function (which is unique for chains with even number of spins) can be represented by an MPS with bond dimension $2$, it is expected that DMRG yields accurate results with a bond dimension cutoff as small as $D=2$.
It should be noted, however, that in this case DMRG takes a few more sweeps to reach convergence.

\section{Conclusion}\label{sec:conclusion}
In summary, we have provided a comprehensive introduction to DMRG, both in the original and in the tensor-network formulations.
For pedagogical purposes, our work is accompanied by concrete practical implementations (see Supplementary Information), the main goal of which is to make the formal description of the method more tangible.
For that reason, our efforts were directed toward producing digestible, transparent and instructive code implementations rather than optimizing their performance or versatility.
Although there exist publicly available user-friendly libraries that efficiently implement DMRG (e.g., TeNPy~\cite{Hauschild2018} and ITensor~\cite{Fishman2022}), we believe that a clear understanding of the method is crucial for an educated use of these resources.
Moreover, it is our opinion that the fundamentals of DMRG are interesting in their own right, as they are at the same time powerful and simple.

Despite not having been covered in this colloquium, extensions of DMRG to tackle quantum dynamics~\cite{Daley2004,Vidal2004,White2004,Schollwock2006} and finite-temperature behavior~\cite{White2009,Stoudenmire2010}---both relevant for the study of out-of-equilibrium quantum many-body phenomena---have been put forth. 
Another topic that was beyond the scope of this review was the exploitation of symmetries~\cite{McCulloch2002,Schollwock2005,Weichselbaum2012} to restrict the DMRG simulations to a given symmetry subspace, both to speed-up the calculations and to find excited states without having to compute and impose the orthogonality with respect to all the lower-energy states.
In any event, upon completing the reading of this manuscript, we are confident the reader is ready to explore the relevant literature to become acquainted with these methods.


\section*{Acknowledgments}
G. C. acknowledges financial support from Funda\c{c}\~{a}o para a Ci\^{e}ncia e a Tecnologia (FCT) for the PhD scholarship grant with reference No. SFRH/BD/138806/2018.
B. M. acknowledges financial support from FCT for the PhD scholarship grant with reference No. SFRH/BD/08444/2020.


\appendix

\section{Gauge freedom of SVD}\label{gauge_SVD}

In the factorization of a matrix via singular value decomposition (SVD), the singular values are unique, but not the singular vectors in general~\cite{Hogben2006}. 

Given a singular value decomposition $U S V^{\dagger}$, there is a gauge freedom associated with the introduction of a resolution of the identity, $\mathds{1} = W W^{\dagger}$, in $U S (W W^{\dagger}) V^{\dagger}$. 
Provided that $W$ commutes with $S$, the matrices $W$ and $W^{\dagger}$ can be absorbed in the definition of the left- and right-singular vectors to produce an alternative singular value decomposition $\tilde{U} S \tilde{V}^{\dagger}$, with $\tilde{U} \equiv U W$ and $\tilde{V}^{\dagger} \equiv W^{\dagger} V^{\dagger}$. 
If all singular values are distinct, then $W$ must also be diagonal to commute with $S$, in which case left- and right-singular vectors are unique up to a phase factor $\mathrm{e}^{\mathrm{i} \theta}$. 
If, instead, there are repeated eigenvalues, then the associated left- and right-singular vectors may be chosen in any fashion such that they span the relevant subspace. 
This corresponds to $W$ being a block-diagonal unitary matrix, with nontrivial blocks associated with the singular vectors of equal singular values. 

It is also possible to introduce on either side of $S$ two resolutions of the identity constructed with a permutation matrix $P$, $U (P^{\dagger} P) S (P^{\dagger} P) V^{\dagger}$, and absorb the permutation matrices as in $U' S' V'^{\dagger}$, with $U' \equiv U P^{\dagger}$, $V'^{\dagger} \equiv P V^{\dagger}$, and $S' \equiv P S P^{\dagger}$. 
The matrix $S'$ is still diagonal, but the order of the entries along the diagonal has changed relative to $S$. 
In particular, this gauge transformation allows to rearrange the singular values in descending order in the definition of $S$. 
This is a common practice, particularly when truncations are considered.

\section{Update of ``effective'' matrix in one-site-update MPS-based DMRG}\label{app:L_R_update}

In this appendix, we explain how to initialize and efficiently update the rank-$3$ $L_{[i]}$ and $R_{[i]}$ tensors that are part of the structure of the ``effective'' matrix $M_{[i]}$ (see Fig.~\ref{fig:Fig_15}) of the eigenvalue problem that results in the optimal update of the local tensor $A_{[i]}$ at site $i \in \{1,2,...,N\}$ of the MPS ansatz (with bond dimension $D$) within the one-site-update finite-system DMRG algorithm.
The Hamiltonian considered has an $N$-site MPO representation with $\mathcal{O}(1)$ bond dimension; its local tensor at site $i$ is denoted by $H_{[i]}$.

Let us assume that the very first site of the MPS to be optimized is the leftmost site, $i = 1$. 
In that case, the initial MPS should be cast in right-canonical form, which takes $\mathcal{O}(N D^3)$ operations.
Before initializing the first left-to-right sweep, a preliminary right-to-left routine (without any local optimization of the MPS) is carried out to compute all $\{ R_{[i]} \}_{i = 1}^{N}$ sequentially.
The initial $R_{[N]}$ is just the $1 \times 1 \times 1$ (reshaped) identity, and then $R_{[N-1]}$ is obtained by contracting the right-normalized $A_{[N]}$, its adjoint $A^{\dagger}_{[N]}$ and $H_{[N]}$ with $R_{[N]}$, following the three-layer closing-the-zipper strategy introduced in Section~\ref{sec:MPO}. 
At the end of this preliminary right-to-left routine, all $\{ R_{[i]} \}_{i = 1}^{N}$ have been computed in $\mathcal{O}(N D^3)$ time. 
Therefore, we see that the initialization of the DMRG algorithm has a computational cost of $\mathcal{O}(N D^3)$.

At this point, all tensors required to define the eigenvalue problem at the first site are available, since the corresponding $L_{[1]}$ operator is the trivial $1 \times 1 \times 1$ identity---there is nothing to the left of site $i = 1$. 
We can therefore start the first left-to-right sweep to optimize the MPS. 
At the end of a given iteration $i$ of this left-to-right sweep, corresponding to the optimization of the local tensor $A_{[i]}$ at site $i$, the rank-$3$ tensor $L_{[i+1]}$ is computed---contracting the previously determined $L_{[i]}$ with the left-normalized $A_{[i]}$, its adjoint $A^\dagger_{[i]}$ and $H_{[i]}$, at $\mathcal{O}(D^3)$ cost---, so that it can be used to define the eigenvalue problem of the next iteration. 
During such left-to-right sweep, the $R_{[i]}$ tensors do not have to be recalculated, because the updated tensors are all absorbed by the $L_{[i]}$ tensors.
Importantly, only a single $L_{[i]}$ tensor is computed at every iteration of the sweep, so the time complexity of one iteration is $\mathcal{O}(D^3)$.

Once a left-to-right sweep is completed and we move on to a right-to-left sweep, the roles are reversed: the $L_{[i]}$ tensors are retrieved from memory and the $R_{[i]}$ tensors are recalculated iteratively.

\section{Trivial truncation of Hilbert space in original version of one-site-update finite-system DMRG}\label{app:no_trunc_original_finite_DMRG}

In the original formulation of the one-site-update finite-system DMRG (see Fig.~\ref{fig:finite-DMRG} but consider only one site, denoted by $\circ$, between blocks S and E), for a $d$-dimensional local degree of freedom and $D$ kept eigenstates, the Hamiltonian of the full system, $\text{S} \circ \text{E}$, is a $(D^2 d) \times (D^2 d)$ matrix, so the target eigenstate $\ket{\psi}$ is a $(D^2 d)$-dimensional vector, which is computed.
Assuming a left-to-right sweep, without loss of generality, the full system in the next iteration, which we denote by $\text{S}' \circ \text{E}'$, has an Hilbert space with increased dimension $D^2 d^2$, since the $D \times D$ Hamiltonian of the shrunk block $\text{E}'$ is fetched from memory, but the Hamiltonian of the grown block $\text{S}' \equiv \text{S} \circ$ is obtained anew, yielding a $(Dd) \times (Dd)$ matrix.
Therefore, the Hilbert space of the block $\text{S}'$ has to be truncated before the diagonalization of the Hamiltonian of $\text{S}' \circ \text{E}'$ takes place.

To compute the reduced density matrices on either side of the bipartition between $\text{S} \circ$ and E, it is useful to obtain the Schmidt decomposition~\cite{Nielsen2010} of $\ket{\psi}$.
This involves reshaping the $(D^2 d)$-dimensional vector $\ket{\psi}$ into a $(D d) \times D$ matrix $M$---according to the considered bipartition---and then performing its full SVD (see Section~\ref{subsec:svd}). 
This yields $M = \mathcal{U} \mathcal{S} \mathcal{V}^{\dagger}$, with $\mathcal{U}$ a $(D d) \times (D d)$ unitary matrix with columns $\{ \ket{u_n} \}_{n = 1}^{D d}$ (the left-singular vectors), $\mathcal{S}$ a $(D d) \times D$ matrix with non-negative real entries along the diagonal (the singular values $\{ s_n \}_{n=1}^D$) and all remaining entries equal to zero, and $\mathcal{V}^\dagger$ a $D \times D$ unitary matrix with lines $\{ \ket{v_n} \}_{n = 1}^{D}$ (the right-singular vectors).

However, as noted in Section~\ref{subsec:svd}, by considering the thin SVD instead, it is possible to convert $\mathcal{S}$, the matrix that encodes the singular values, into a $D \times D$ matrix $S$ by discarding the corresponding $(D d - D)$ columns of $\mathcal{U}$, resulting in the left-normalized $(D d) \times D$ matrix $U$. 
These discarded columns are nothing more than left-singular vectors associated with zero-valued rows of $\mathcal{S}$, so this truncation is exact.

In the end, the Schmidt decomposition reads as
\begin{equation}
    \ket{\psi} = \sum_{n = 1}^{D} s_n \ket{u_n}_{\text{S} \circ} \otimes \ket{v_{n}}_{\text{E}},
\end{equation}
and the reduced density matrices on either side can be written as
\begin{equation}
    \begin{aligned}
    & \sigma_{\text{S} \circ} = \textrm{Tr}_{\text{E}}(\ket{\psi} \bra{\psi}) = \sum_{n = 1}^{D} s_{n}^{2} \ket{u_{n}}_{\text{S} \circ} \prescript{}{\text{S} \circ}{\bra{u_{n}}}, \\
    & \sigma_{\text{E}} = \textrm{Tr}_{\text{S} \circ}(\ket{\psi} \bra{\psi}) = \sum_{n = 1}^{D} s_{n}^{2} \ket{v_{n}}_{\text{E}} 
    \prescript{}{\text{E}}{\bra{v_{n}}}.
    \end{aligned}
\end{equation}

In summary, we have shown that the eigenvalues of the reduced density matrices are the square of the singular values obtained by performing the SVD of the target eigenstate in the corresponding bipartition, thus establishing a connection between the original and the MPS-based formulations of DMRG.
Moreover, we have found that $\sigma_{\text{S} \circ}$, which is generally a $(Dd) \times (Dd)$ matrix, only has $D$ eigenvectors with nonzero eigenvalues, which can be used to truncate the Hilbert space of the block $\text{S}'$ without any approximation, thus showing why no actual truncation takes place in the original formulation of the one-site update finite-system DMRG algorithm, as in the corresponding MPS-based version.

\bibliography{mybib}

\end{document}